\documentclass[useAMS,usenatbib]{mn2e}

\def\hi{\relax \ifmmode {\mbox H\,{\scshape i}}\else H\,{\scshape i}\fi}
\def\hii{\relax \ifmmode {\mbox H\,{\scshape ii}}\else H\,{\scshape ii}\fi}
\def\hei{\relax \ifmmode {\mbox He\,{\scshape i}}\else He\,{\scshape i}\fi}
\def\heii{\relax \ifmmode {\mbox He\,{\scshape ii}}\else He\,{\scshape ii}\fi}
\def\nii{\relax \ifmmode {\mbox N\,{\scshape ii}}\else N\,{\scshape ii}\fi}
\def\ni{\relax \ifmmode {\mbox N\,{\scshape i}}\else N\,{\scshape i}\fi}
\def\oi{\relax \ifmmode {\mbox O\,{\scshape i}}\else O\,{\scshape i}\fi}
\def\oii{\relax \ifmmode {\mbox O\,{\scshape ii}}\else O\,{\scshape ii}\fi}
\def\oiii{\relax \ifmmode {\mbox O\,{\scshape iii}}\else O\,{\scshape iii}\fi}
\def\sii{\relax \ifmmode {\mbox S\,{\scshape ii}}\else S\,{\scshape ii}\fi}
\def\siii{\relax \ifmmode {\mbox S\,{\scshape iii}}\else S\,{\scshape iii}\fi}
\def\ariii{\relax \ifmmode {\mbox Ar\,{\scshape iii}}\else Ar\,{\scshape iii}\fi}
\def\ariv{\relax \ifmmode {\mbox Ar\,{\scshape iv}}\else Ar\,{\scshape iv}\fi}
\def\neiii{\relax \ifmmode {\mbox Ne\,{\scshape iii}}\else Ne\,{\scshape iii}\fi}
\def\cliii{\relax \ifmmode {\mbox Cl\,{\scshape iii}}\else Cl\,{\scshape iii}\fi}
\def\feiii{\relax \ifmmode {\mbox Fe\,{\scshape iii}}\else Fe\,{\scshape iii}\fi}
\def\feii{\relax \ifmmode {\mbox Fe\,{\scshape ii}}\else Fe\,{\scshape ii}\fi}
\def\niqii{\relax \ifmmode {\mbox Ni\,{\scshape ii}}\else Ni\,{\scshape ii}\fi}
\def\cii{\relax \ifmmode {\mbox C\,{\scshape ii}}\else C\,{\scshape ii}\fi}
\def\mgi{\relax \ifmmode {\mbox Mg\,{\scshape i}}\else Mg\,{\scshape i}\fi}
\def\silii{\relax \ifmmode {\mbox Si\,{\scshape ii}}\else Si\,{\scshape ii}\fi}
\def\ha{\relax \ifmmode {\mbox H}\alpha\else H$\alpha$\fi}
\def\hb{\relax \ifmmode {\mbox H}\beta\else H$\beta$\fi}
\def\me{$^{-1}$}
\def\arcsec{\hbox{$^{\prime\prime}$}}
\def\arcmin{\hbox{$^{\prime}$}}
\def\deg{\hbox{$^{\circ}$}}
\usepackage{array}
\usepackage{graphicx}
\usepackage{lscape}
\usepackage{epsfig,natbib}
\usepackage{longtable}
\usepackage{amssymb}
\usepackage{changepage}
\title[Chemical abundances in M31 from H II regions]{Chemical abundances in M31 from H II regions}
\author[A. Zurita and F. Bresolin]{A. Zurita$^{1,2}$\thanks{E-mail:
azurita@ugr.es} and F. Bresolin$^{3}$\footnotemark[1]\thanks{E-mail:bresolin@ifa.hawaii.edu}\\
$^{1}$Dpto. de F\'\i sica y del Cosmos, Campus de Fuentenueva, Edificio Mecenas, Universidad de Granada, 18071--Granada, Spain
\\
$^{2}$Instituto Carlos I de F\'\i sica Te\'orica y Computacional, Facultad de Ciencias, 18071--Granada, Spain\\
$^{3}$Institute for Astronomy, 2680 Woodlawn Drive, Honolulu, HI 96822, USA}
\begin{document}

\date{}

\pagerange{\pageref{firstpage}--\pageref{lastpage}} \pubyear{2012}

\maketitle

\label{firstpage}

\begin{abstract}
 We have obtained multi--slit spectroscopic observations from 3700\,\AA\ to 9200\,\AA\ with LRIS at the Keck I telescope for 31 \hii\ regions 
 in the disk of the Andromeda galaxy (M31), spanning a range in galactocentric distance from 3.9~kpc to 16.1~kpc. In 9 \hii\ regions we measure one 
 or several auroral lines ([\oiii]$\lambda4363$,  [\nii]$\lambda5755$, [\siii]$\lambda6312$, [\oii]$\lambda7325$), from which we determine 
 the electron temperature ($T_e$) of the gas and  derive chemical abundances using the {\em direct $T_e$--based method}. We analyze, for the first time in
 M31, abundance trends with galactocentric radius from the {\em direct} method, and find that the Ne/O, Ar/O, N/O  and  
 S/O abundance ratios are consistent with a constant value across the M31 disc, while the O/H abundance ratio shows a weak gradient.
 We have combined our data with all spectroscopic observations of \hii\ regions in M31 available in the literature, yielding 
 a sample of 85 \hii\ regions spanning distances from 3.9~kpc to 24.7~kpc (0.19\,--\,1.2~R$_{25}$) from the galaxy center.  We have tested  a number of 
 empirical calibrations of strong emission line ratios. We find that the slope of the oxygen abundance gradient in M31 is $-0.023\pm0.002$~dex~kpc\me, and that the central oxygen abundance is in the range 12+log(O/H)\,=\,8.71\,--\,8.91~dex  (i.e.~between 1.05 and 1.66 times the solar value, 
for  12+log(O/H)$_\odot$=8.69), depending on the calibration adopted.
 The \hii\ region oxygen abundances  are compared  with the results from other metallicity indicators (supergiant stars and 
 planetary nebulae). The comparison shows that \hii\ region O/H abundances are systematically $\sim$0.3 dex below the stellar ones.  This discrepancy is  
 discussed in terms of oxygen depletion onto dust grains and possible biases affecting $T_e$--based oxygen abundances at high metallicity.
 \end{abstract}


\section{Introduction}
M31, the Andromeda galaxy, is our nearest neighbour and the largest galaxy of 
the Local Group and these properties make  this galaxy a fundamental benchmark 
for understanding spirals and their evolution. M31's proximity allows for spatially detailed studies of its constituent stellar 
populations, even though the large angular extent on the sky hampers a global view of this galaxy.
The advent of large field cameras and multi--object spectrographs has been critical 
for a growing amount of information on important properties of M31.
These improvements in observational capabilities have permitted, for example, the discovery of a giant stellar tidal 
stream in the halo \citep{ibata2001} and detailed studies of the resolved stellar 
populations both in the disc and in the halo 
\citep[e.g.][]{ferguson2002,ferguson2005,davidge,tanaka2010} which have radically 
changed our knowledge on the formation and evolution of M31.
It is  now recognized that interactions and satellite accretions 
have been widely responsible for the evolution of the Andromeda galaxy 
\citep[e.g.][]{ibata2001,bernard2012,davidge,hammer2007}.
The interstellar medium also shows signatures of disruptions possibly resulting through
these interactions  \citep[e.g.][]{unwin,corbelli}. The current star formation 
rate (SFR) across the disc, $\sim$1~M$_\odot$~yr\me\ \citep{williams}, is significantly lower than in the Milky Way
 and most of the star--forming sites are
distributed in a prominent ring of approximately $\sim$10~kpc galactocentric radius.
This morphology has been predicted with dynamical models of the recent interaction of M31 
with a companion galaxy \citep[possibly M32,][]{block_nature}.

The chemical evolution of M31 has also been extensively studied 
in comparison with the Milky Way (\citealt*{marcon-uchida}; \citealt{yin09}; \citealt{renda05}).
Both Local Group galaxies seem to share common features in their chemical evolution,
but in order to explain the observations in detail,  models require 
different formation histories for the two galaxies. In particular, M31 must have 
been more active and efficient in the past in forming stars than 
the Milky Way \citep{yin09,renda05}.

The present--day chemical element abundances and their spatial distribution across a galaxy are key 
observational properties for constraining  different models of chemical evolution 
\citep[e.g.][]{marcon-uchida}.
\hii\ regions have been traditionally the main targets used to trace the present-day elemental abundances
across galaxy disks. However, nebular metallicity data on M31 are surprisingly sparse. Since the first 
\hii\ region emission--line strength measurements made by \citet{rubin} in the early 1970's, in which
the existence of an abundance gradient was suggested,
only a handful of authors have tried to measure the abundance gradient of M31. 
\citet{dennefeld81} analyze a sample of 12 supernova remnants (SNR) and 8 \hii\ regions. They 
report gradients in the N/H and N/O ratios across the galaxy disk, but no gradient in the
distribution of oxygen abundance with galactocentric radius, within the large uncertainties in
their abundances determinations.
In a subsequent study, \citet*{Blair82} obtain an oxygen abundance gradient with a slope of approximately $-$0.027~dex~kpc\me
and a super--solar oxygen abundance in the M31 center (12+log(O/H)~$>$~9.0) 
from the application of empirical {\em strong--line} nebular metallicity determination methods 
to the observed line fluxes in a sample of 11 \hii\ regions. In the same paper they calculate the oxygen abundance 
gradient from the spectra of 9 SNRs, and these show no significant gradient across the disk, with
a mean abundance 12+log(O/H)\,$\simeq$\,8.5. 
The most recent spectroscopic observations of a sample of \hii\ regions in M31 date from 
the late 1990's and were presented by \citet*{Galarza} and \citet*{B99}. The former use the  
R$_{23}$ parameter and find a somewhat steeper oxygen gradient, with a slope of $-$0.06$\pm$0.03~dex~kpc\me.
All these authors derive the chemical abundance gradient in M31 by applying empirical 
calibrations of bright--line ratios, as the temperature sensitive auroral lines used to constrain
electron temperature and emissivities of abundance-sensitive forbidden lines (i.e.~through the {\em direct method}) 
remained undetected. In fact, it is somewhat surprising that, to-date, there are no auroral line detections for the determination of 
\hii\ region abundances through the {\em direct method} in M31, apart from a single \hii\ region 
(K932) observed by \citet{esteban09}.

Other attempts to measure the oxygen abundance gradient in M31 come  from the
reanalysis of the  datasets  mentioned above (mainly those of the 1980's) and 
the application of different {\em strong-line methods} (\citealt*{zaritsky94};
\citealt{vila-costas}; \citealt{smartt}; \citealt{carrie}).
The M31 oxygen abundance gradient  slope derived by the different authors ranges from 
$-0.013$~dex~kpc\me\ to $-0.06$~dex~kpc\me, i.e. they differ by as much
as a factor of four, mainly due to  uncertainties and differences in the  
strong--line methods used. The central oxygen abundance that results from 
these studies ranges between 1 and 3.5 times the solar value.

Motivated by the desirability of a modern re-evaluation of the present--day chemical abundance  in 
M31, we secured new spectra of \hii\ regions in 
the Andromeda galaxy, in order to obtain nebular chemical abundances based on the detection of 
the faint auroral lines used for electron temperature determinations. 
The main  aim of the current paper is to measure the oxygen abundance gradient from \hii\ regions in M31 and to 
compare with other metallicity indicators, in particular young, massive stars \citep{Venn,carrie,smartt} and planetary nebulae \citep{kwitter}.

This paper is organized as follows. The spectroscopic observations,
the data reduction procedure and the methodology employed for meassuring emission line fluxes 
are described in Sec.~\ref{observations_reduction}. In Sec.~\ref{Te} we derive the physical 
properties (electron densities and temperatures) of the \hii\ regions and from these, the ionic 
and total chemical abundances are obtained (Sec.~\ref{abundances}). The radial oxygen abundance 
gradient is then calculated (Sec.~\ref{O_grad})
from our {\em direct measurements} and from empirical metallicity calibrations applied to the \hii\ regions of our sample and 
to a compilation of data from other authors. A comparison with
other metallicity indicators is presented in Sec.~\ref{comparison}. Finally,  in Sec.~\ref{discussion} with discuss 
and summarize our results.

%{\bf Notas importantes para lo que queda:
%SFR in MW ~3-6 Msun/yr (Boissier & Prantzos 1999) comparado con ~0.3-1.0 en M31 (Barmby et al. 2006; Williams 2003)

\section{Observations and data reduction}
\label{observations_reduction}
\subsection{Observations}

\begin{table}
\begin{minipage}{\textwidth}
\caption{Journal of observations.}
\label{observations}
\begin{tabular}{l c c c}    
\hline
Mask  &  $\Delta\lambda$  &  Spectral range\footnote{Approximate spectral range for a slit in the centre of the detector.} & Exposure Time \\
          &    (\AA , FWHM)         &     (\AA)              & (s) \\
\hline
 1 & 5.6  & 3450--6000  & 3$\times$1800\,+\,2$\times$1200 \\ 
 1 & 4.5  & 5120--6840 & 3$\times$1800 \\        
 1 & 9.3  & 6450--10000 & 2$\times$1200 \\        
 2 & 5.6  & 3450--6000  & 3$\times$1800\,+\,2$\times$1000 \\        
 2 & 4.5  & 5120--6840 & 3$\times$1800 \\         
 2 & 9.3  & 6450--10000 & 2$\times$1000 \\         
 3 & 5.6  & 3450--6000  & 2$\times$1800\,+\,2$\times$1000 \\        
 3 & 4.5  & 5120--6840 & 2$\times$1800  \\    
 3 & 9.3  & 6450--10000 & 2$\times$1000  \\    
\hline
\end{tabular}
\end{minipage}
\end{table}
The spectroscopic observations of \hii\ regions in the disk of M31 were 
carried out on 2005 September 29 with the Low Resolution Imaging 
Spectrometer \citep[LRIS,][]{Oke95}  at the 10m Keck I telescope operated 
by the W.M. Keck Observatory on the Mauna Kea summit.
The sky was photometric with seeing $\sim$0.8'' at the beginning 
of the night. Conditions slowly got worse, having seeing 
$\sim$1.5'' at the end of the night.
\begin{table}
\begin{minipage}{\textwidth}
\caption{M31 adopted parameters.}
\label{M31 params}
\begin{tabular}{l c l }    
\hline
Parameter &  Value &  References  \\
\hline
RA (J2000)     & 00h 42m 44.33s      &  \citet{2mass}\\ 
Dec (J2000)    & +41\deg 16' 07.50'' &  \citet{2mass}\\
Inclination    &  77\deg                  &  \citet{corbelli}\\
Position angle &  38\deg                  &  \citet{corbelli}\\
Distance           & (744\,$\pm$\,33)~kpc  &  \citet{vilardell10}\\
R$_{25}$   (arcmin) & (95\,$\pm$\,2) & \citet{rc3}\\
R$_{25}$    (kpc)   & (21\,$\pm$\,1) & \\
\hline
\end{tabular}
\end{minipage}
\end{table}

\begin{figure*}
   \centering
   \includegraphics[bb= 7 6 855 608, clip, width=\textwidth]{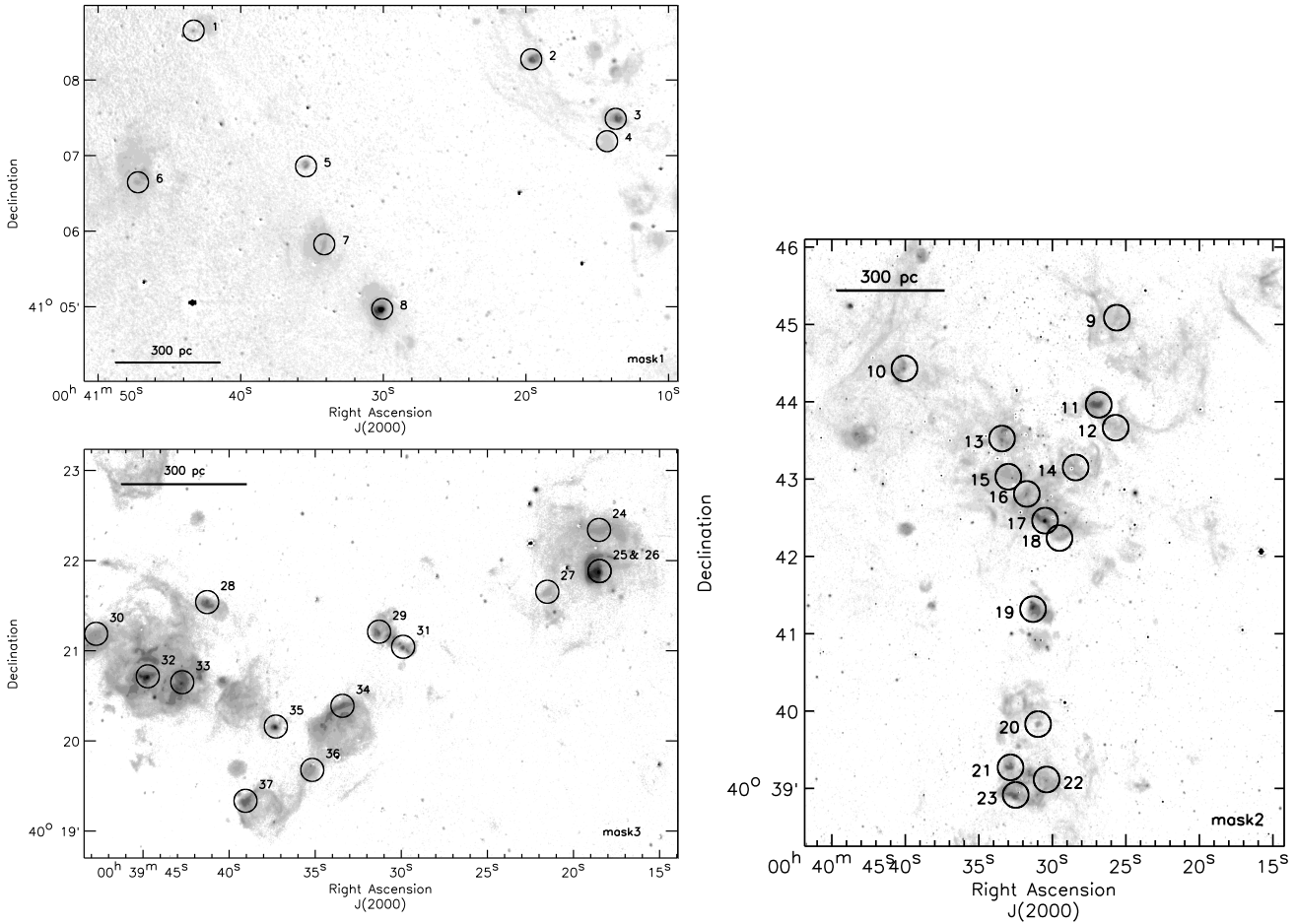}
   \caption{Identification of the targets included in our LRIS Keck multi-slit masks plotted over an \ha\ continuum-subtracted image obtained at the 
    2.5m Isaac Newton Telescope at la Palma. The circles mark the brightest \ha\ knots along each slit, where the apertures for the one dimensional 
    spectrum extractions were centred.}
\label{masks}
\end{figure*}
LRIS was used in multi-slit mode. Both the blue and red channels of the spectrograph were 
used simultaneously with a dichroic beam splitter (with 50\% transmittance at 5091\,\AA) for two 
different setups. The first setup was used to cover the spectral range 3500\,\AA\ to 6840\,\AA\ for 
a slit at  the center of the detector with spectral resolution and
dispersion $\sim$5.6\,\AA\  (FWHM) and 0.62\,\AA/pix, respectively, in the blue channel (with a 
600~lines~mm$^{-1}$ grism blazed at 4000\,\AA) and $\sim$4.5\,\AA\ (FWHM) and 0.62\,\AA/pix in 
the red channel (with a  900~lines~mm$^{-1}$ grating blazed at 5500\,\AA). The second 
setup was identical in the blue channel, but included a 400~lines~mm$^{-1}$ grating blazed
at 8500\,\AA\ in the red arm, covering roughly from 6450\,\AA\ to 10000\,\AA\  
with 9.3\,\AA\ FWHM resolution. 
\begin{figure*}
   \centering
   \includegraphics[width=15.5cm]{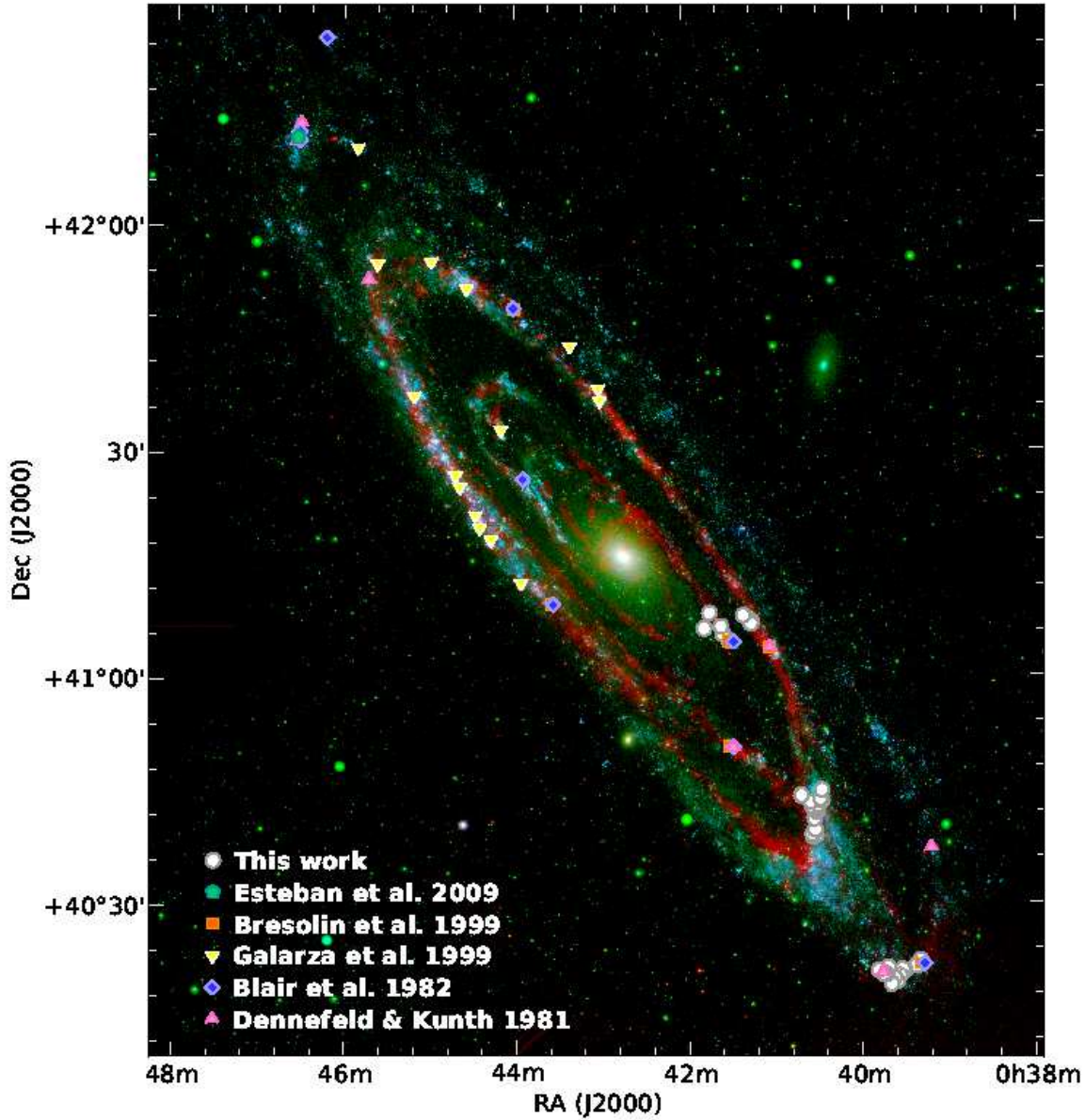}
   \caption{Image of M31 showing the location of the \hii\ studied in this work 
   (white circles).  \hii\ regions studied by other authors are indicated with pink triangles \citep{dennefeld81}, violet diamonds \citep{Blair82}, yellow squares
    \citep[showing slit centers by][]{Galarza}, inverted orange triangles \citep{B99} and cyan pentagon \citep{esteban09}. The image is a composite of GALEX 
   (FUV: blue, NUV: green) and Spitzer (24 $\mu$m: red).}
   \label{our_regions}
\end{figure*}
The blue channel spectra were imaged on a mosaic of two 2K$\times$4K Marconi (E2V) CCDs giving 
a plate scale of $0.135''$~pix$^{-1}$, the red channel spectra on a back-illuminated Tektronix
2K$\times$2K CCD, with $0.215''$~pix$^{-1}$.

Three slit masks were defined and used for the multi--object spectroscopy, centered
approximately 17\arcmin, 42\arcmin\/ and 66\arcmin\ from the nucleus of M31.
The  masks cover  
4.8\arcmin$\times$7.3\arcmin\ on the sky,
and contained 8, 15 and 13 1.2\arcsec-wide slits, respectively. The slit lengths varied between
7\arcsec\ and 1.5\arcmin.
The observations were carried out ensuring that the slit position angle in each mask was
within 20\deg\ of the parallactic angle for airmasses greater than 1.3, in order to  minimize light 
losses due to differential atmospheric refraction.
Table~\ref{observations} contains a journal of the observations.

The target \hii\ regions were selected from   \ha\ continuum-subtracted images  (see Fig.~\ref{masks}), previously obtained
with the Wide Field Camera of the 2.5m Isaac Newton Telescope (INT) at La Palma
(Spain) within the Local Group Census project\footnote{The Local Group Census is a 
 narrow-band imaging survey of all the galaxies of the Local Group above Dec\,=\,$-30$\deg 
 (http://www.ing.iac.es/$\sim$rcorradi/LGC/).} \citep[e.g.][]{LGC}. An astrometric calibration of these images
with an accuracy of 0.3 arcsec was secured using stellar positions from the USNO2  \citep{USNO2} catalogue.
%allowed us to carry out the mask design with the {\tt autoslit3} software.
The positions of the selected \hii\ regions across the M31 disc are shown in Fig.~\ref{our_regions} and their
coordinates are given in Table~\ref{Galactocentric distances}.
Four targets are in common with the \cite{Blair82} sample (BA423, BA379, BA416, BA370; the last two were classified as 
SNRs by these authors).
\subsection{Data reduction}
The data reduction was carried out using standard IRAF\footnote{IRAF 
is distributed by the National Optical Astronomy Observatories, which 
is operated by the Association of Universities for Research in Astronomy, 
Inc. (AURA) under cooperative agreement with the National Science Foundation.}
techniques. 
After removing the bias level, a correction was applied to account for the 
different gains (e$^{-}$/ADU) of the amplifiers of the red and blue CCDs.
Afterwards the spectrum corresponding to each slit was identified 
and separated out from the 2--dimensional slit mask spectra of science and 
calibration frames, and in subsequent tasks the spectrum from each slit 
was treated as a separate spectrum. The data reduction was then
carried out using standard procedures for long--slit spectra: flatfielding, cosmic-ray
removal \citep[{\tt lacos$\_$spec},][]{lacos}, removal of 
geometrical distortions in the spectral and spatial directions, wavelength calibration, 
extraction of one dimensional spectra within
defined apertures and combination of the individual spectra.

\begin{table*}
\centering
\begin{minipage}{\textwidth}
\centering
\caption{\hii\ region equatorial coordinates and galactocentric distances.}
\label{Galactocentric distances}
\begin{tabular}{l c c c c c c }     
\hline
 ID\footnote{Nebulae with auroral line detections are indicated with $^{\dag}$. A double dagger ($^\ddag$) indicates supernova remnant candidates.}       &    RA    $ \, \, \, \, \, \, \, \, \, \, \, \, \, \, \, \, \, \,$  Dec  
& Galactocentric distance\footnote{Galactocentric distances computed from the M31 orientation parameters shown in Table~\ref{M31 params}.}  &Other ID\footnote{IDs starting with {\em  BA} from \citet{BA64} 
and with {\em P} from \citet{pellet}. Nebulae  marked with $\ast$ have been spectroscopically observed by \citet{Blair82}.} \\
                 &           (J2000)                 &              (kpc)     &   \\         
\hline
   1                      &00:41:43.67 +41:08:40.27    & 5.05     &  -- \\
   2$^{\dag}$               &00:41:19.94 +41:08:21.38    &8.16       &   BA 419, P 318\\ 
   3$^{\dag}$              & 00:41:13.96 +41:07:35.09    &8.64      &   BA 422, P 310\\ 
   4                      & 00:41:14.54 +41:07:17.25   & 8.42     &   P 308\\ 
   5$^{\dag,}$$^{\ddag}$     &00:41:35.65  +41:06:53.86    & 5.42      &  BA 416 (SNR candidate), P 325$^{\ast}$  \\  % BA416$^{**}$ (classified as SNR) \\  
   6                     & 00:41:47.42 +41:06:39.13    &  3.93     &  BA 415 (SNR candidate), P 327  \\
   7                     &00:41:34.30 +41:05:52.14     & 5.23      &   P 317 \\
   8$^{\dag,}$$^{\ddag}$     &00:41:30.17  +41:05:01.51    & 5.47     &  BA 423, P 309$^{\ast}$ \\ % BA423$^{**}$\\  
   9                      & 00:40:26.10 +40:45:08.45   & 8.85      & P 182\\ 
  10                      & 00:40:40.52 +40:44:26.75   & 8.58      & --\\  
  11                     & 00:40:27.26 +40:44:00.71    & 8.93      &  BA 464, P 176\\  
  12                      & 00:40:26.08 +40:43:43.06   & 9.01      & P 173\\  
  13                      & 00:40:33.81 +40:43:33.39   &8.86      & P 179\\  
  14$^{\ddag}$             & 00:40:28.77 +40:43:11.74    & 9.02     & P 171\\  
  15                      & 00:40:33.35 +40:43:03.86   & 8.97      & P 174?\\  
  16                      & 00:40:32.07 +40:42:50.72   &9.04      & P 174?\\  
  17$^{\dag}$              &00:40:30.81 +40:42:30.19     & 9.13      & BA 310, P 167\\
  18                      & 00:40:29.81 +40:42:16.84   & 9.19       & --\\  
  19$^{\dag}$              & 00:40:31.55 +40:41:21.40    &9.41       & BA 312, P 161\\
  20                      & 00:40:31.11 +40:39:52.24    & 9.85       & --\\ 
  21                      & 00:40:32.94 +40:39:18.35    & 10.07      & BA 318, P 150\\  
  22$^{\ddag}$             & 00:40:30.48 +40:39:09.33     &10.07        & P 145\\  
  23                      & 00:40:32.57 +40:38:57.02    &10.18       & BA 319, P 149\\  
  24                      &00:39:18.84 +40:22:26.04    & 14.53       &BA 375, P 15 \\  
  25$^{\dag}$              &00:39:18.77 +40:21:58.66     &14.66       & BA 379, P 7$^{\ast}$ \\%BA379$^{**}$ \\
  26$^{\dag}$              & 00:39:18.77 +40:21:58.66    & 14.66      & BA 379, P 7$^{\ast}$ \\%BA379$^{**}$ \\
  27                      & 00:39:21.79 +40:21:44.31    & 14.74     &BA 378, P 9 \\ 
  28                      & 00:39:41.55 +40:21:33.54    & 15.25      &BA 358, P 32 \\  
  29$^{\dag,}$$^{\ddag}$     & 00:39:31.53 +40:21:15.81   &15.06         & BA 368, P 21 \\%BA370$^{**}$ (classified as SNR) \\
  30                      & 00:39:47.97 +40:21:11.07   & 15.67       &BA 357, P 34 \\  
  31                      & 00:39:30.12 +40:21:06.05   & 15.09       & BA 373\\  
  32$^{\dag,}$$^{\ddag}$     &00:39:44.93 +40:20:43.43     &15.73          & BA 360\\
  33$^{\dag}$               &00:39:42.92 +40:20:39.90    & 15.67        & BA 362\\
  34$^{\dag}$               & 00:39:33.59 +40:20:26.06  & 15.42         &BA 374, P 17 \\                                       
  35$^{\dag}$               & 00:39:37.44 +40:20:11.49    & 15.65      &  BA 371, P 20\\ 
  36                      & 00:39:35.28 +40:19:43.06   & 15.76       &BA 377, P 11 \\  
  37                      & 00:39:39.13 +40:19:21.71   & 16.06       &BA 376, P 14 \\  
\hline                                                    
\end{tabular}                                             
\end{minipage}
\end{table*}

The final spectra were flux calibrated using observations of the spectrophotometric standards BD+28~4211, 
G191-B2B, Hiltner~600 and GD71 that were obtained at the beginning, middle and end of the night with a 
8.7\arcsec-wide slit. The standard star fluxes have been obtained from \citet{oke90}, \citet*{bohlin95},
and \citet{hamuy94}. A standard mean optical/near-IR extinction curve for Mauna Kea was adopted 
for the atmospheric extinction  correction (from the  UKIRT web 
pages\footnote{http://www.jach.hawaii.edu/UKIRT/astronomy/utils/exts.html. This extinction curve 
comes from CFHT Bulletin \citep*{mkoextinct88} and the CFHT observer's manual.}).

Given the limited near-IR range of most of the spectrophotometric standard star spectra, only data 
below 9200\,\AA\ are useful. The [\siii]$\lambda$9532 fluxes have been calculated from the theoretical 
ratio [\siii]$\lambda$9532/[\siii]$\lambda$9069=2.44.

\subsection{Emission line flux measurements and interstellar extinction correction}
\label{line_fluxes}
The emission line fluxes of the  \hii\ regions were measured with the {\tt SPLOT} IRAF task, by 
integrating all the flux in the emission line between two given limits over the local continuum level
defined interactively.

We analysed the spectra of 37 nebulae. Two of them correspond to different extractions of
the same slit, in which  two bright knots of ionized gas emission (objects 25 and 26 in Table~\ref{Galactocentric distances}) were identified.
For 13 targets we were able to detect one or several of the following auroral 
lines, [\oiii]$\lambda4363$, [\nii]$\lambda5755$, [\siii]$\lambda6312$,  and [\oii]$\lambda7325$, which 
allow the determination of the electron temperature of the 
different ions, although four of the targets turned out to be SNR, as described below.
The other 24 targets have  lower signal--to--noise spectra and no auroral lines were detected. 
We  identified two SNR candidates within this subsample.
For the remaining 22 \hii\ regions we concentrate here only on the brightest emission lines, which will be used for the determination 
of the oxygen abundances through {\em strong-line methods} (Sect.~\ref{strong}).

The three different spectral ranges  (blue, red and near--infrared) were scaled in order to bring them
to the same relative flux calibration. Several 
emission lines in common between the red and the near-infrared spectral ranges (typically  \ha, [\nii]$\lambda\lambda$6548,\,6583,
 He{\sc I}~$\lambda$6678, [\sii]$\lambda\lambda$6717,\,6731) provided the 
scaling factor. There are no lines in common between the blue and
the red spectral ranges. In this case a preliminary scaling factor was obtained from the continuum level of both 
spectra. The final scaling factor was later determined by imposing the expected  \ha/\hb\ flux ratio \citep{storey95} 
for the measured or assumed electron temperature, after the interstellar extinction and stellar 
absorption corrections had been applied (see below).

The interstellar reddening was determined from the  Balmer series hydrogen recombination 
lines flux decrement. As the fluxes of nebular Balmer emission lines are affected by
absorption by the underlying stellar population, a correction needs to be applied in order to
obtain the interstellar extinction  to properly correct our measurements. 
We have assumed that the equivalent width of the stellar absorption (EW$_{abs}$) is the same for 
all Balmer lines. Then, an iterative procedure was performed: the extinction coefficient c(\hb) 
was determined from the comparison of the measured fluxes of the H$\gamma$ and H$\delta$ lines
relative to H$\beta$ and the expected ratios from the case B recombination given by \citet{storey95} for an
electron temperature of $7500$~K and density of 100~cm$^{-3}$. The \citet{howarth83} parametrization  of the 
interstellar reddening law by \citet{seaton79} was adopted for R$_V$=3.1. The EW$_{abs}$ was then 
adjusted in successive iterations until the c(\hb) value obtained from the 
  H$\gamma$/\hb\ and H$\delta$/\hb\ line ratios were in agreement. 
After the determination of the electron temperature (see Sect.~\ref{Te}) the 
iterative process was repeated. We checked that higher order recombination lines of the 
Balmer series measured in our spectra were in agreement (within errors) with the theoretical
predictions.
The extinction coefficient c(\hb) obtained varies between 0.24 and 1.0 in our \hii\ region sample, while the EW$_{abs}$ was found 
to be in the range 1.5--3.2\,\AA\ for the highest S/N spectra, where we detected the 
auroral lines.
For the lowest S/N spectra, the extinction coefficient c(\hb) was typically determined from 
the H$\gamma$/\hb\ ratio alone (and in this case the EW$_{abs}$ was assumed to be zero) or from both
H$\gamma$/\hb\ and  H$\delta$/\hb\ when possible. c(\hb) for these regions ranges from 
0.0 to 1.03.

The final values of c(\hb), EW$_{abs}$, and the blue--red scaling factor were further checked 
in the highest S/N \hii\ region spectra  by  comparing the observed \hei\ line ratios with 
theoretical predictions \citep*{benjamin99}. 
 The observed and theoretical \hei\ line ratios 
$\lambda$5876/$\lambda$4471, $\lambda$6678/$\lambda$4471 and $\lambda$5876/$\lambda$4922 
differ by less than 15\% in all targets, except for \#5, for which differences are larger (of order $\sim$40\%).
We note that the observational errors for these ratios are also higher ($\sim$30\%) for this target, 
which turned out to be a SNR (see Sect.~\ref{SNRs}).
The same check was done with the Balmer ratio H$\epsilon$/\hb\ and  
Paschen series line ratios to \hb\ when available (using mainly Pa 9, Pa 10, Pa 11, Pa 12).
 In this case the observed and theoretical line ratios differ by less than 20\% (typically $\sim$5-10\%), 
except for the H$\epsilon$/\hb\ ratio in targets \#5, \#25, \#26 and \#35 for which the difference 
observed--theoretical is larger, $\sim$25-30\%, except for target \#5 (SNR) in which H$\epsilon$ is twice the value expected 
for an  \hii\ region possibly due to blending with enhanced [\neiii] \citep*{stupar}.

Table~\ref{lines} shows the line fluxes corrected for reddening and 
normalized to \hb\ for the 13 nebulae where auroral lines fluxes have been measured. For the 
remaining targets (with no detections of auroral lines) we give in  Table~\ref{bl_line_fluxes} the
brightest line fluxes corrected for reddening and 
normalized to \hb. The coordinates and galactocentric distances for both sets of regions are given 
in  Table~\ref{Galactocentric distances}.

The  uncertainties quoted in Tables~\ref{lines} and \ref{bl_line_fluxes} include (added in quadrature): 
statistical errors measured with {\tt SPLOT}, 
flatfielding errors, the uncertainty in the flux calibration, and the 
uncertainty in the determination of the flux scaling factors between the different 
spectral ranges and in the determination of the extinction coefficient. Typically, errors in
the line flux ratios are of order $\sim$4-8\% for lines brighter than F$_{H\beta}$/2, and  
$\sim$6-16\% for fainter lines, for the \hii\ regions with detections of auroral lines.
For the regions with lower S/N  the errors in the line flux ratios relevant for {\em strong-line
methods} for abundance determinations are of order $\sim$5-20\% for [\oii]/\hb, $\sim$5-40\% for [\oiii]$\lambda4959$/\hb,
$\sim$4-30\% for [\oiii]$\lambda5007$/\hb, and $\sim$5-20\% for [\nii]$\lambda6583$/\hb.

%Typical errors are:
% flatfielding: ~1%
% flux calibration: ~3%
% scaling factors: 2-4%
% extinction constant: 

Four of our targets are located within $\sim$3\arcsec\ of the positions reported 
by \citet[see Table~\ref{Galactocentric distances}]{Blair82}. We have compared the fluxes published by these authors with the 
fluxes of our brightest lines. The measurements agree within 15\% for lines brighter than 
F$_{H\beta}$/3 across the whole spectral range in common  (from 3700\,\AA\ to 6730\,\AA).
The differences are  higher for  [\sii]$\lambda\lambda$6717,\,6731 in \#5 (BA416),  
$\sim$35\%, and  for [\oii]$\lambda$3727 in \#25 (BA379), $\sim$50\%. 
For line ratios fainter than F$_{H\beta}$/3 the differences increase up to $\sim$60\% (in both directions), 
with no dependence on wavelength. We note that \citet{Blair82} state that
their error in faint line fluxes ($<$\,0.5\,F$_{H\beta}$) may be as large as 50\%.
\subsection{Supernova remnants}
\label{SNRs}
Targets  \#5 (BA416) and  \#29 (BA370)  are classified as SNRs by \citet{Blair82}. 
For both nebulae we find [\sii]/\ha\,$>$\,0.5, indicating the presence of shocks. Our measured line ratios 
for BA416  position this region in the
 area of the  \ha/[\sii] vs. \ha/[\nii] diagnostic diagram that corresponds to SNR 
 \citep*{sabbadin}, in agreement with the conclusion by \citet{Blair82}. According to our measurements, the location of  BA370  in the same
diagnostic diagram does not allow us to clearly establish the nature of this nebula. However, we measure a substantial
[\oi]$\lambda$$\lambda$6300,\,6360 flux ([\oi]/\hb\,$>$\,0.4). Following the line ratio diagnostic criteria of \cite*{fesen}
([\oii]/\hb\ vs. [\oi]/\hb) both BA416 and BA370 are identified as SNRs. We note that, as mentioned above, \citet{Blair82}
reported a $\sim$35\% higher   [\sii]$\lambda\lambda$6717,\,6731 flux for BA416 than  determined from our observations. Given that
the center position of our slit does not perfectly coincide with theirs, we cannot exclude some dilution of the   
SNR emission by surrounding photoionized gas emission in our spectrum.

%%%%%%%%%%%%%%%%%%%%%%%%%%%%%%%%%%%%%%%%%%%%%%%%%%%%%%%%%%%%%%%%%%%%%%%%%%%%%%%%%%%%%%%%%%%%%%%%%%%%%%%%%%%%%%%%%%%%%%%%%%%%%%%%%%%%%%%%%%%%%%%%%%%%%%%%%%%%%%%%%%%%%%%%%%%%%%%%%%%%%%%%%%%%%%%%%%%%%%%%%%%%%%%%%%%%%%
%---------------Tables with extinction--corrected line ratios
%\clearpage 
%\newpage
\onecolumn
\changetext{3.5em}{1.0em}{}{}{}
\begin{landscape}
\begin{flushleft}
%\hspace{-10cm}
%\begin{longtable}{l l c c c c c c c c c c c c c}
%\begin{longtable}{>{\tiny}l >{\tiny}l  >{\tiny}c  >{\tiny}c >{\tiny}c >{\tiny}c >{\tiny}c >{\tiny}c >{\tiny}c  >{\tiny}c  >{\tiny}c  >{\tiny}c  >{\tiny}c  >{\tiny}c  >{\tiny}c }  
%\begin{longtable}{>{\footnotesize}l >{\footnotesize}l  >{\footnotesize}c  >{\footnotesize}c >{\footnotesize}c >{\footnotesize}c >{\footnotesize}c >{\footnotesize}c >{\footnotesize}c  >{\footnotesize}c  >{\footnotesize}c  >{\footnotesize}c  >{\footnotesize}c  >{\footnotesize}c  >{\footnotesize}c }  
\begin{longtable}{>{\small}l >{\small}l  >{\small}c  >{\small}c >{\small}c >{\small}c >{\small}c >{\small}c >{\small}c  >{\small}c  >{\small}c  >{\small}c  >{\small}c  >{\small}c  >{\small}c }  
\caption{Reddening--corrected line fluxes in nebulae with auroral line detections.}
\label{lines}\\
\hline
  Line& $\lambda_\circ$      &   \#2    &\#3   &  \#5$^{*}$&    \#8$^{*}$   &      \#17      & \#19      &\#25      &\#26  & \#29$^{*}$  &\#35       & \#32$^{*}$       &\# 33      & \#34 \\  
            &(\,\AA)          &         &      &          &               &                &            &         &      &             &           &                 &           &      \\
\hline
\endfirsthead
\caption{Continued.}\\
         
\hline
  Line &$\lambda_\circ$ &  \#2             &\#3       &  \#5$^{*}$&       \#8$^{*}$   &      \#17      &      \#19          &\#25             &\#26        & \#29$^{*}$        &\#35           & \#32$^{*}$       &\# 33          & \#34 \\ 
       &(\,\AA)          &         &      &          &               &                &            &         &      &             &           &                 &           &      \\
\hline
\endhead
\hline
\endfoot                                                                                                                        
H 16+\hei\  &3704  &        ...     &  1.34$\pm$0.08 &     ...     &  0.98$\pm$0.05 &  1.11$\pm$0.10 &        ...     &  1.00$\pm$0.06 &  1.42$\pm$0.12  &     ...       &         ...    &         ...    &         ...    &         ...     \\ 
H 15        &3712  &        ...     &  1.35$\pm$0.08 &     ...     &  0.85$\pm$0.04 &  1.48$\pm$0.13 &        ...     &  1.07$\pm$0.07 &  0.96$\pm$0.08  &     ...       &         ...    &         ...    &         ...    &         ...     \\ 
$[$\oii$]$  &3727  &  210$\pm$20    &   170$\pm$10   &  770$\pm$60 &   150$\pm$7    &   143$\pm$13   &     159$\pm$8  &    158$\pm$10  &   230$\pm$20    & 330$\pm$20    &  150$\pm$10    &   269$\pm$20   &  220$\pm$2     &   289$\pm$14    \\ 
H 12        &3750  &  1.94$\pm$0.14 &  2.39$\pm$0.14 &     ...     &  1.69$\pm$0.08 &  1.9$\pm$0.2   &        ...     &  1.87$\pm$0.12 &  2.9$\pm$0.2    &     ...       &  3.1$\pm$0.2   &  1.73$\pm$0.10 &  1.57$\pm$0.11 &         ...     \\
H 11        &3771  &   5.4$\pm$0.3  &  5.2$\pm$0.2   &     ...     &  5.0$\pm$0.2   &  4.0$\pm$0.3   &        ...     &  4.9$\pm$0.3   &  4.5$\pm$0.3    & 6.0$\pm$0.3   &  4.3$\pm$0.2   &  4.0$\pm$0.2   &  7.3$\pm$0.4   &  4.7$\pm$0.2    \\
H 10        &3798  &   7.0$\pm$0.4  &  6.0$\pm$0.3   & 4.1$\pm$0.3 &  6.0$\pm$0.2   &  5.3$\pm$0.4   &  7.0$\pm$0.3   &  5.8$\pm$0.3   &  5.8$\pm$0.5    & 7.3$\pm$0.4   &  5.1$\pm$0.3   &  5.3$\pm$0.3   &  6.6$\pm$0.4   &  7.5$\pm$0.3    \\
\hei\       &3819  &        ...     &        ...     &     ...     &  0.52$\pm$0.02 &  0.69$\pm$0.06 &        ...     &         ...    &  1.19$\pm$0.10  &     ...       &  0.90$\pm$0.06 &  0.73$\pm$0.04 &         ...    &         ...     \\ 
H 9         &3835  &  9.0$\pm$0.5   &  8.5$\pm$0.4   &10.4$\pm$0.8 &  8.0$\pm$0.3   &  6.9$\pm$0.6   &  8.5$\pm$0.3   &  7.8$\pm$0.4   &  7.9$\pm$0.6    & 8.0$\pm$0.4   &  7.0$\pm$0.4   &  7.2$\pm$0.4   &  8.0$\pm$0.5   &  7.0$\pm$0.3    \\ 
$[$\neiii$]$&3868  &   2.2$\pm$0.2  &        ...     &  77$\pm$6   &  0.71$\pm$0.03 &  4.4$\pm$0.4   &        ...     & 14.7$\pm$0.9   & 13.2$\pm$1.1    &     ...       & 18.7$\pm$1.2   &  3.8$\pm$0.2   &         ...    &  5.8$\pm$0.3    \\ 
H 8         &3889  &  15.3$\pm$1.1  & 14.9$\pm$0.9   &18.3$\pm$1.4 & 13.0$\pm$0.6   & 16.2$\pm$1.4   & 13.7$\pm$0.7   & 16.4$\pm$1.0   & 19$\pm$2        & 15.7$\pm$0.9  & 17.6$\pm$1.1   & 15.1$\pm$0.8   & 12.4$\pm$0.9   & 17.0$\pm$0.8    \\     
H$\epsilon$ &3969  &  16.4$\pm$1.1  & 15.1$\pm$0.8   &  32$\pm$2   & 12.4$\pm$0.5   & 16.4$\pm$1.3   & 14.1$\pm$0.6   & 19.8$\pm$1.2   & 20$\pm$2        &15.1$\pm$0.8   & 21.1$\pm$1.3   & 16.2$\pm$0.8   & 15.7$\pm$1.0   & 16.0$\pm$0.7    \\     
\hei\       &4026  &  0.78$\pm$0.05 &        ...     &     ...     &  0.85$\pm$0.04 &  1.62$\pm$0.13 &        ...     &  1.05$\pm$0.06 &  1.68$\pm$0.14  &     ...       &  2.04$\pm$0.12 &  1.09$\pm$0.06 &         ...    &         ...     \\
%$[$\sii$]$  &4069  &  1.22$\pm$0.08 &        ...     &12.9$\pm$0.9 &  2.2$\pm$0.10  &  0.94$\pm$0.08 &  0.96$\pm$0.05 &         ...    &         ...     &     ...       &         ...    &  2.8$\pm$0.2   &         ...    &         ...     \\
%$[$\sii$]$  &4076  &  0.54$\pm$0.04 &        ...     & 5.5$\pm$0.4 &  0.73$\pm$0.03 &  0.30$\pm$0.02 &  1.00$\pm$0.05 &         ...    &         ...     &     ...       &         ...    &  0.76$\pm$0.04 &         ...    &         ...     \\
$[$\sii$]$  &4072  &  1.75$\pm$0.12 &        ...     &18.4$\pm$1.3 &  2.90$\pm$0.14 &  1.2$\pm$0.1   &  1.94$\pm$0.09 &  0.94$\pm$0.06 &  1.30$\pm$0.11  & 4.9$\pm$0.3   &  1.00$\pm$0.06 &  3.5$\pm$0.2   &         ...    &         ...     \\
H$\delta$   &4101  &     25$\pm$2   & 25.5$\pm$1.4   &  27$\pm$2   & 25.3$\pm$1.1   &     25$\pm$2   &  25.5$\pm$1.1  &     25$\pm$2   & 26$\pm$2        &25.5$\pm$1.4   & 25$\pm$2       & 25.3$\pm$1.3   & 25.5$\pm$2     & 25.9$\pm$1.2    \\     
 \hei\      &4144  &        ...     &        ...     &     ...     &        ...     &        ...     &        ...     &         ...    &  0.47$\pm$0.04  &     ...       &         ...    &         ...    &         ...    &         ...     \\     
$[$\feii$]$ &4244  &        ...     &        ...     &     ...     &  0.20$\pm$0.01 &        ...     &        ...     &         ...    &         ...     &               &         ...    &         ...    &         ...    &         ...     \\     
\cii\       &4267  &        ...     &        ...     &     ...     &  0.22$\pm$0.01 &        ...     &        ...     &         ...    &         ...     &     ...       &         ...    &         ...    &         ...    &         ...     \\     
$[$\feii$]$ &4287  &        ...     &        ...     &     ...     &  0.23$\pm$0.01 &        ...     &        ...     &         ...    &         ...     &     ...       &         ...    &         ...    &         ...    &         ...     \\     
H$\gamma$   &4340  &     46$\pm$3   &    47$\pm$3    &  44$\pm$3   &    46$\pm$2    &    47$\pm$3    &    46$\pm$2    &      47$\pm$3  & 47$\pm$4        &47$\pm$3       & 47$\pm$3       & 47$\pm$2       & 47$\pm$3       & 47$\pm$2        \\     
$[$\oiii$]$ &4363  &        ...     &        ...     &22.1$\pm$1.4 &  0.34$\pm$0.02 &  0.40$\pm$0.03 &        ...     &  1.30$\pm$0.08 &  1.20$\pm$0.10  &     ...       &  1.35$\pm$0.08 &  0.97$\pm$0.05 &         ...    &         ...     \\     
\hei\       &4388  &        ...     &        ...     &     ...     &  0.27$\pm$0.01 &  0.41$\pm$0.03 &        ...     &  0.23$\pm$0.01 &  0.61$\pm$0.05  &     ...       &  0.81$\pm$0.05 &  0.33$\pm$0.02 &         ...    &         ...     \\     
$[$\feii$]$ &4414  &        ...     &        ...     & 3.7$\pm$0.2 &  0.36$\pm$0.02 &        ...     &        ...     &         ...    &         ...     &     ...       &         ...    &  0.53$\pm$0.03 &          ...   &         ...     \\     
\hei\       &4471  &  2.9$\pm$0.2   &  3.4$\pm$0.2   &2.19$\pm$0.13&  2.82$\pm$0.13 &  3.8$\pm$0.3   &  2.23$\pm$0.11 &  3.6$\pm$0.2   &  4.2$\pm$0.3    & 2.41$\pm$0.14 &  4.3$\pm$0.3   &  3.2$\pm$0.2   &  2.2$\pm$0.2   &  4.3$\pm$0.2  \\     
\mgi$]$     &4563  &        ...     &        ...     &     ...     &  0.24$\pm$0.01 &        ...     &        ...     &         ...    &  0.15$\pm$0.01  &     ...       &         ...    &  0.37$\pm$0.02 &         ...    &         ...     \\     
\mgi$]$     &4571  &        ...     &        ...     &     ...     &  0.16$\pm$0.01 &        ...     &        ...     &         ...    &         ...     &     ...       &         ...    &  0.15$\pm$0.01 &         ...    &         ...     \\     
$[$\feiii$]$&4658  &        ...     &        ...     & 4.6$\pm$0.3 &  0.89$\pm$0.04 &        ...     &        ...     &         ...    &         ...     &     ...       &         ...    &  0.99$\pm$0.05 &         ...    &         ...     \\     
\heii\      &4686  &        ...     &        ...     & 7.4$\pm$0.4 &  1.57$\pm$0.07 &        ...     &        ...     &         ...    &         ...     &     ...       &         ...    &         ...    &         ...    &         ...     \\     
$[$\feiii$]$&4702  &        ...     &        ...     &2.23$\pm$0.13&  0.23$\pm$0.01 &        ...     &        ...     &         ...    &  0.08$\pm$0.01  &     ...       &         ...    &  0.33$\pm$0.02 &         ...    &         ...     \\     
\hei\       &4713  &        ...     &        ...     & 3.2$\pm$0.2 &  0.24$\pm$0.01 &  0.37$\pm$0.02 &        ...     &  0.31$\pm$0.02 &  0.45$\pm$0.04  &     ...       &         ...    &  0.32$\pm$0.02 &         ...    &         ...     \\     
$[$\ariv$]$ &4740  &       ...      &        ...     &1.36$\pm$0.08&       ...      &        ...     &        ...     &       ...      &      ...        &     ...       &         ...    &         ...    &         ...    &          ...   \\
$[$\feiii$]$&4755  &        ...     &        ...     &2.02$\pm$0.12&  0.16$\pm$0.01 &        ...     &        ...     &         ...    &         ...     &     ...       &         ...    &         ...    &         ...    &         ...     \\     
H$\beta$    &4861  &  100$\pm$7     &  100$\pm$6     & 100$\pm$6   &  100$\pm$5     &  100$\pm$7     &100$\pm$5       &  100$\pm$6     &  100$\pm$8      &  100$\pm$6    &   100$\pm$6    &    100$\pm$5   & 100$\pm$7      &  100$\pm$5      \\     
 $[$\feiii$]$&4881  &        ...     &        ...     &1.35$\pm$0.08&  0.27$\pm$0.01 &        ...     &        ...     &         ...    &         ...     &     ...       &         ...    &  0.25$\pm$0.01 &         ...    &         ...     \\     
\hei\       &4922  &  0.77$\pm$0.05 &  0.91$\pm$0.05 &0.70$\pm$0.04&  0.80$\pm$0.04 &  1.04$\pm$0.07 &        ...     &  0.91$\pm$0.05 &  1.17$\pm$0.09  &     ...       &  1.22$\pm$0.07 &  0.76$\pm$0.04 &         ...    &         ...     \\     
$[$\oiii$]$ &4959  &     22$\pm$2   & 15.6$\pm$0.9   &122$\pm$7    &   8.9$\pm$0.4  &    40$\pm$3    & 11.7$\pm$0.6   &    86$\pm$5    &    75$\pm$6     &12$\pm$0.7     &104$\pm$6       & 36$\pm$2       & 20.5$\pm$1.4   & 59$\pm$2        \\     
$[$\feiii$]$&4986  &        ...     &        ...     &5.1$\pm$0.3  &  0.57$\pm$0.03 &        ...     &        ...     &  0.16$\pm$0.01 &  0.24$\pm$0.02  &     ...       &         ...    &  1.26$\pm$0.06 &         ...    &         ...     \\     
$[$\oiii$]$ &5007  &     58$\pm$4   & 44$\pm$2       & 350$\pm$20  &    27$\pm$1    &   114$\pm$8    &    34$\pm$2    &  251$\pm$14    &  220$\pm$20     &34$\pm$2       & 320$\pm$20     &  102$\pm$5     & 59$\pm$4       &   180$\pm$9     \\     
\hei\       &5016  &  1.62$\pm$0.11 &  1.82$\pm$0.10 &     ...     &  1.69$\pm$0.08 &  2.2$\pm$0.2   &  1.49$\pm$0.07 &  2.09$\pm$0.12 &  2.2$\pm$0.2    &     ...       &  2.8$\pm$0.2   &  1.84$\pm$0.09 &  1.51$\pm$0.10 &  2.03$\pm$0.10  \\     
$[$\feii$]$ &5159  &       ...      &        ...     &3.4$\pm$1.1  &       ...      &        ...     &        ...     &       ...      &      ...        &     ...       &         ...    &         ...    &         ...    &          ...    \\
$[$\ni$]$   &5199  &        ...     &  0.75$\pm$0.08 &   18$\pm$2  &   3.4$\pm$0.2  &  0.43$\pm$0.07 &  0.9$\pm$0.2   &  0.56$\pm$0.05 &  0.91$\pm$0.08  & 8.5$\pm$0.6   &  0.57$\pm$0.14 &         ...    &         ...    &  1.3$\pm$0.2   \\     
$[$\feii$]$ &5262  &        ...     &        ...     & 1.7$\pm$1.0 &  0.11$\pm$0.03 &        ...     &        ...     &         ...    &         ...     &     ...       &         ...    &         ...    &         ...    &         ...     \\     
$[$\feiii$]$&5270  &        ...     &        ...     &   5$\pm$2   &  0.46$\pm$0.03 &        ...     &        ...     &         ...    &         ...     &     ...       &         ...    &  0.56$\pm$0.04 &         ...    &         ...     \\     
$[$\cliii$]$&5518  &  0.25$\pm$0.05 &  0.30$\pm$0.04 &     ...     &  0.27$\pm$0.02 &  0.35$\pm$0.05 &        ...     &  0.46$\pm$0.04 &  0.39$\pm$0.05  &     ...       &  0.40$\pm$0.09 &  0.33$\pm$0.03 &         ...    &         ...     \\     
$[$\cliii$]$&5538  &  0.19$\pm$0.04 &  0.20$\pm$0.04 &     ...     &  0.23$\pm$0.02 &  0.24$\pm$0.04 &        ...     &  0.30$\pm$0.03 &  0.29$\pm$0.05  &     ...       &  0.25$\pm$0.07 &  0.24$\pm$0.03 &         ...    &         ...     \\     
$[$\nii$]$  &5755  &  0.50$\pm$0.08 &  0.69$\pm$0.06 & 6.1$\pm$1.1 &  0.86$\pm$0.05 &  0.43$\pm$0.06 &  0.51$\pm$0.10 &  0.40$\pm$0.03 &  0.66$\pm$0.06  & 1.3$\pm$0.2   &  0.45$\pm$0.08 &  0.78$\pm$0.05 &  0.70$\pm$0.07 &  1.6$\pm$0.2    \\     
\hei\       &5876  &   9.8$\pm$1.0  & 10.5$\pm$0.6   &10.6$\pm$1.3 &   9.2$\pm$0.4  &  12.9$\pm$1.0  &  9.4$\pm$0.5   & 11.6$\pm$0.7   & 11.8$\pm$0.8    & 7.1$\pm$0.5   & 11.4$\pm$0.9   &  9.1$\pm$0.5   &  7.2$\pm$0.4   & 11.7$\pm$0.6    \\     
\oi         &5959  &        ...     &        ...     &     ...     &  0.09$\pm$0.02 &        ...     &        ...     &         ...    &         ...     &     ...       &         ...    &         ...    &         ...    &         ...     \\     
\silii\     &5978  &        ...     &        ...     &     ...     &  0.13$\pm$0.02 &        ...     &        ...     &         ...    &         ...     &     ...       &         ...    &  0.11$\pm$0.02 &         ...    &         ...     \\     
$[$\oi$]$   &6300  &  1.24$\pm$0.14 &        ...     &73$\pm$6     &   5.6$\pm$0.3  &  0.88$\pm$0.09 &  1.40$\pm$0.14 &  1.12$\pm$0.07 &  1.77$\pm$0.13  &27$\pm$2       &  0.94$\pm$0.10 &  6.0$\pm$0.4   &  1.60$\pm$0.12 &  5.6$\pm$0.3    \\     
$[$\siii$]$ &6312  &  0.60$\pm$0.07 &        ...     & 2.6$\pm$0.8 &  0.60$\pm$0.04 &  0.97$\pm$0.11 &        ...     &  1.21$\pm$0.08 &  1.37$\pm$0.10  &     ...       &  1.45$\pm$0.14 &         ...    &          ...   &         ...     \\     
\silii\     &6347  &        ...     &        ...     &     ...     &  0.25$\pm$0.03 &        ...     &        ...     &         ...    &         ...     &     ...       &         ...    &         ...    &         ...    &         ...     \\     
$[$\oi$]$   &6360  &  0.39$\pm$0.06 &        ...     &23$\pm$2     &  2.11$\pm$0.11 &  0.27$\pm$0.04 &        ...     &  0.28$\pm$0.03 &  0.59$\pm$0.05  & 9.8$\pm$0.6   &         ...    &  2.01$\pm$0.13 &  0.38$\pm$0.05 &  1.7$\pm$0.2    \\     
$[$\nii$]$  &6548  &    26$\pm$3    & 26.7$\pm$2     &  121$\pm$11 &    44$\pm$2    &    20$\pm$2    &     32$\pm$2   & 10.7$\pm$0.7   & 16.3$\pm$1.1    &33$\pm$2       & 11.0$\pm$0.9   & 21.1$\pm$1.3   & 24$\pm$2       & 32$\pm$2       \\     
H$\alpha$   &6563  &   290$\pm$30   &  300$\pm$20    &  300$\pm$30 &   293$\pm$20   &   289$\pm$3    &    300$\pm$20  &290$\pm$20      &  280$\pm$20     &  294$\pm$20   & 300$\pm$30     &  280$\pm$20    &   290$\pm$20   &  284$\pm$14    \\     
$[$\nii$]$  &6583  &    81$\pm$8    &    80$\pm$4    &  370$\pm$30 &   133$\pm$7    &    63$\pm$6    &    97$\pm$5    &    32$\pm$2    &    48$\pm$3     &  101$\pm$6    &    34$\pm$3    &   64$\pm$4     & 71$\pm$5       &   98$\pm$5     \\     
\hei\       &6678  &   2.9$\pm$0.3  &  3.2$\pm$0.2   & 3.2$\pm$0.9 &  2.59$\pm$0.13 &   3.5$\pm$0.3  &   2.7$\pm$0.2  &  3.3$\pm$0.2   &  3.4$\pm$0.3    & 2.3$\pm$0.2   &  3.4$\pm$0.3   &  2.6$\pm$0.2   &  2.2$\pm$0.2   &  2.9$\pm$0.2   \\     
$[$\sii$]$  &6717  &    23$\pm$2    & 22.5$\pm$1.2   &  160$\pm$20 &    44$\pm$2    &    15$\pm$2    &     29$\pm$2   & 10.9$\pm$0.7   & 17.7$\pm$1.3    &94$\pm$6       & 12.7$\pm$1.1   & 39$\pm$3       & 28$\pm$2       & 42$\pm$2       \\     
$[$\sii$]$  &6731  &    17$\pm$2    & 16.5$\pm$0.9   & 131$\pm$13  &    36$\pm$2    & 11.0$\pm$1.1   & 20.1$\pm$1.1   &  7.9$\pm$0.6   & 12.4$\pm$0.9    &70$\pm$5       &  9.2$\pm$0.8   & 29$\pm$2       & 19.7$\pm$1.3   & 29$\pm$2       \\     
\hei\       &7065  &   1.3$\pm$0.2  &  1.39$\pm$0.11 &     ...     &  1.19$\pm$0.07 &  1.9$\pm$0.2   &  1.3$\pm$0.3   &  2.0$\pm$0.8   &  1.9$\pm$0.2    & 2.1$\pm$0.6   &  2.0$\pm$0.2   &  1.3$\pm$0.8   &  1.5$\pm$0.2   &  1.6$\pm$0.3   \\     
$[$\ariii$]$&7135  &   6.1$\pm$0.7  &  6.0$\pm$0.4   & 10$\pm$4    &   4.6$\pm$0.2  &  8.9$\pm$1.0   &  5.5$\pm$0.5   &   9.8$\pm$0.9  &  9.7$\pm$0.7    & 3.1$\pm$0.5   & 10.8$\pm$1.0   &  6.1$\pm$1.0   &  5.4$\pm$0.4   & 11.7$\pm$0.7   \\     
\cii\       &7236  &        ...     &        ...     & 2.2$\pm$0.5 &        ...     &  0.35$\pm$0.06 &        ...     &         ...    &         ...     &     ...       &         ...    &         ...    &         ...    &         ...    \\     
\hei\       &7281  &   0.5$\pm$0.2  &        ...     &     ...     &        ...     &  0.68$\pm$0.09 &        ...     &         ...    &         ...     &     ...       &         ...    &         ...    &         ...    &         ...   \\     
$[$\oii$]$  &7325  &   2.0$\pm$0.3  &        ...     &37$\pm$5     &  3.0$\pm$0.2   &  2.4$\pm$0.3   &        ...     &  3.2$\pm$0.8   &  5.0$\pm$0.4    &     ...       &         ...    &  6.1$\pm$1.1   &  6.2$\pm$0.5   &  8.9$\pm$0.6  \\     
$[$\niqii$]$&7378  &       ...      &        ...     & 8.6$\pm$1.1 &       ...      &        ...     &        ...     &       ...      &      ...        &     ...       &         ...    &         ...    &         ...    &          ...   \\
$[$\ariii$]$&7751  &   1.3$\pm$0.2  &        ...     &     ...     &  1.04$\pm$0.06 &  2.2$\pm$0.3   &        ...     &  2.3$\pm$0.5   &  2.2$\pm$0.2    &     ...       &  2.5$\pm$0.4   &  1.5$\pm$0.7   &         ...    &         ...     \\     
Pa 16       &8502  &   0.7$\pm$0.2  &        ...     &     ...     &        ...     &        ...     &        ...     &         ...    &         ...     &     ...       &         ...    &         ...    &         ...    &         ...     \\     
Pa 15       &8545  &   1.0$\pm$0.2  &        ...     &     ...     &        ...     &        ...     &        ...     &         ...    &         ...     &     ...       &         ...    &         ...    &         ...    &         ...     \\     
Pa 14       &8598  &   1.3$\pm$0.2  &        ...     &     ...     &        ...     &        ...     &        ...     &         ...    &         ...     &     ...       &         ...    &         ...    &         ...    &         ...     \\     
Pa 13       &8665  &   3.7$\pm$0.5  &        ...     &     ...     &        ...     &        ...     &        ...     &         ...    &         ...     &     ...       &         ...    &         ...    &         ...    &         ...     \\     
Pa 12       &8750  &   1.1$\pm$0.2  &        ...     &     ...     &  1.21$\pm$0.06 &        ...     &        ...     &         ...    &         ...     &     ...       &         ...    &  0.94$\pm$0.14 &         ...    &         ...     \\     
Pa 11       &8863  &   1.4$\pm$0.2  &        ...     &     ...     &  1.43$\pm$0.07 &        ...     &        ...     &         ...    &         ...     &     ...       &         ...    &         ...    &         ...    &         ...     \\     
Pa 10       &9015  &   1.7$\pm$0.3  &        ...     &     ...     &  1.60$\pm$0.08 &  2.1$\pm$0.3   &        ...     &         ...    &         ...     &     ...       &         ...    &         ...    &         ...    &  ... \\     
$[$\siii$]$ &9069  &    23$\pm$3    &  20.8$\pm$1.3  &11$\pm$2     &    19$\pm$1    &   23$\pm$3     & 19.3$\pm$1.3   & 21$\pm$2       & 22$\pm$2        & 8.00$\pm$2    & 21$\pm$2       & 16$\pm$2       & 20$\pm$3       & 13.4$\pm$1.2  \\     
Pa 9        &9229  &   2.5$\pm$0.4  &        ...     &  ...        &  2.43$\pm$0.13 &  2.9$\pm$0.4   &        ...     &  2.4$\pm$0.2   &  2.0$\pm$0.2    &     ...       &  2.3$\pm$0.4   &  2.0$\pm$0.8   &  2$\pm$2       &         ...     \\   
\multicolumn{2}{l}{C(H$\beta$)}     &  0.21$\pm$0.04 & 0.31$\pm$0.02  &0.76$\pm$0.1 &  0.67$\pm$0.02 & 0.55$\pm$0.10  &0.44$\pm$0.02   & 0.43$\pm$0.05  &  0.42$\pm$0.04  & 0.34$\pm$0.04 &1.00$\pm$0.05   &  0.39$\pm$0.05 &0.54$\pm$0.04   & 0.44$\pm$0.02\\ 
\multicolumn{2}{l}{EW(H$\beta$)}    & 189.6$\pm$0.4  & 198.7$\pm$0.4  &166.6$\pm$1.1& 156.90$\pm$0.14&226.1$\pm$0.4   &74.9$\pm$0.3    & 123.4$\pm$0.1  & 377.6$\pm$0.6   & 174.1$\pm$0.8 &  908$\pm$2     & 186.0$\pm$0.2  &82.3$\pm$0.2    & 184.0$\pm$0.5\\  
\multicolumn{2}{l}{F(H$\beta$)$^{**}$}    & 11.1$\pm$0.5   & 13.0$\pm$0.5   &3.9$\pm$0.2  & 142$\pm$5      &21$\pm$1        &9.8$\pm$0.3     & 53$\pm$2       & 25.5$\pm$1.4    & 3.5$\pm$0.1    &26$\pm$1       & 35.7$\pm$1.2   &17.3$\pm$0.8    & 9.4$\pm$0.3\\ 
\hline	
%\end{tabular}
%\end{minipage}
\end{longtable}
{\footnotesize \begin{tabular}{l}
%Notes: \\
$^{*}$ indicates SNRs candidates (see Sect. \ref{line_fluxes}) for details. \\
$^{**}$ F(H$\beta$) is the total extinction--corrected \hb\ flux in units of $10^{-15}$ erg cm$^{-2}$ s\me.\\
\end{tabular}}
\end{flushleft}
\end{landscape}
\changetext{-3.5em}{-1.0em}{}{}{}
\twocolumn
%%%%%%%%%%%%%%%%%%%%%%%%%%%%%%%%%%%%%%%%%%%%%%%%%%%%%%%%%%%%%%%%%%%%%%%%%%%%%%%%%%%%%%%%%%%%%%%%%%%%%%%%%%%%%%%%%%%%%%%%%%%%%%%%%
Four additional targets (\#8, \#14, \#22 and \#32) are SNR candidates. Targets \#8 and \#32 have a 
[\sii]/\ha\ ratio consistent with an \hii\ region, but their spectra present numerous  
metal lines in the blue range (mainly from Fe and Mg), as well as relatively strong 
[\oi]$\lambda$$\lambda$6300,\,6360.  In addition, their [\oiii]$\lambda$4363 emission appears to be 
contaminated by [\feii]$\lambda$4359. We suspect that these targets are young SNRs.

For  targets with lower signal--to--noise spectra, we have tentatively identified as SNRs those  
with [\sii]/\ha\,$\gtrsim$\,0.4  and substantial [\oi]$\lambda$6300 emission, 
[\oi]$\lambda$6300\,$\gtrsim$\,0.02, \citep{dennefeld81}. According to this criteria, targets 
\#14 and  \#22 are SNR candidates as well. 

The six targets discussed in this section are removed from our sample and will not be considered in the remainder
of the paper, since the application of \hii\ region metallicity diagnostics would not be appropriate. For completeness 
we report their line fluxes  in Table~\ref{lines} (\#5,  \#8,  \#29 and \#32) and in Table~\ref{bl_line_fluxes}  (\#14 and \#22).

%%%%%%%%%%%%%%%%%%%%%%%%%%%%%%%%%%%%%%%%%%%%%%%%%%%%%%%%%%%%%%%%%%%%%%%%%%%%%%%%%%%%%%%%%%%%%%%%%%%%%%%%%%%%%%%%%%%%%%%%%%%%%%%%%
\begin{table*}
\centering
\begin{minipage}{\textwidth}
\caption{Reddening--corrected line fluxes relative to \hb\ (with \hb=100) for the \hii\ regions with no detections of auroral lines.}
\label{bl_line_fluxes}
\begin{tabular}{l c c c c c c c c c c}     % 11 columns
\hline	
   ID     & [\oii]    & [\oiii] &[\oiii] &  [\nii] & \ha\  &[\nii]& [\sii]     &[\sii]  & F(\hb)                      &c(\hb) \\
          & 3727      & 4959      &5007  &6548     &6562 &6583& 6717       &6731    &(10$^{-15}$ erg s\me cm$^{-2}$) &(mag) \\
\hline
1        &   ....       &   ...       &  ...          &44\,$\pm$\,2    &290\,$\pm$\,20 &149\,$\pm$\,8  &...          &...        &  0.20\,$\pm$\,0.08  & 0.00  \\
4        &  140\,$\pm$\,30  &  ...        &   ...         &20\,$\pm$\,4    &290\,$\pm$\,60 &60\,$\pm$\,13  & 27\,$\pm$\, 6   & 19\,$\pm$\,4    &0.5\,$\pm$\,0.1    & 0.64 \\ 
6        &  122\,$\pm$\,6   &4.7\,$\pm$\,0.5  & 11.6\,$\pm$\,0.7  &33\,$\pm$\,2    &300\,$\pm$\,20 &122\,$\pm$\,7  & 40\,$\pm$\,2    & 25\,$\pm$\,2    &1.43\,$\pm$\,0.05  & 0.00 \\ 
7        &  230\,$\pm$\,20  &2.1\,$\pm$\,0.9  &  3.5\,$\pm$\,1.0  &51\,$\pm$\,5    &290\,$\pm$\,30 &161\,$\pm$\,15 & ...         &...          &1.09\,$\pm$\,0.04  & 0.45 \\ 
 9       &  200\,$\pm$\,20 &12\,$\pm$\,2     & 35\,$\pm$\,3      &38\,$\pm$\,4    &290\,$\pm$\,30 &123\,$\pm$\,12 & 77\,$\pm$\,8    & 53\,$\pm$\,5    &0.42\,$\pm$\,0.02  &0.00 \\ 
10       & 150\,$\pm$\,20  & ...         & 7\,$\pm$\,2       &45\,$\pm$\,6    &290\,$\pm$\,40 &130\,$\pm$\,16 & 48\,$\pm$\,6    & 33\,$\pm$\,4    &1.00\,$\pm$\,0.04  &0.58 \\ 
11       &  150\,$\pm$\,10 & 10.3\,$\pm$\,0.6& 29.5\,$\pm$\,1.4  &31\,$\pm$\,2    &290\,$\pm$\,20 &93\,$\pm$\,7   & 30\,$\pm$\,2    & 21\,$\pm$\,2    &3.9\,$\pm$\,0.1    &0.19 \\  
12       & 149\,$\pm$\,12  &6.2\,$\pm$\,1.4  & 18\,$\pm$\,2      & 40\,$\pm$\,4   &290\,$\pm$\,30 &120\,$\pm$\,11 & 72\,$\pm$\,7    & 46\,$\pm$\,5    &0.232\,$\pm$\,0.009 &0.00 \\  
13       &   97\,$\pm$\,8  &6.5\,$\pm$\,0.8  & 21\,$\pm$\,1      &26\,$\pm$\,2    &290\,$\pm$\,30 &75\,$\pm$\,7   &...           &...         &1.15\,$\pm$\,0.04   &0.36 \\  
14$^{*}$ & 390\,$\pm$\,50 &...          & 16\,$\pm$\,3      &29\,$\pm$\,6    &280\,$\pm$\,50 &100\,$\pm$\,20 & 90\,$\pm$\,20   & 69\,$\pm$\,13   &0.56\,$\pm$\,0.04   &0.59 \\  
15       & 180\,$\pm$\,20  &12\,$\pm$\,4     & 27\,$\pm$\,4      &30\,$\pm$\,5    &280\,$\pm$\,50 &80\,$\pm$\,13  & 47\,$\pm$\,8    & 29\,$\pm$\,5    &0.68\,$\pm$\,0.03   &0.64 \\  
16       &   94\,$\pm$\,8  &  ...        & 4.6\,$\pm$\,0.8   &18\,$\pm$\,2    &290\,$\pm$\,30 &67\,$\pm$\,6   & 36\,$\pm$\,4    & 24\,$\pm$\,2    &0.58\,$\pm$\,0.02   &0.00 \\  
18       & 193\,$\pm$\,20  & ...         & ...           &39\,$\pm$\,4    &300\,$\pm$\,30 &124\,$\pm$\,12 & 72\,$\pm$\,7    & 50\,$\pm$\,5    &0.68\,$\pm$\,0.03   &0.68 \\ 
20       & 130\,$\pm$\,8   &6.6\,$\pm$\,0.6  & 18\,$\pm$\,1      &25\,$\pm$\,2    &290\,$\pm$\,20 &73\,$\pm$\,5   & 31\,$\pm$\,2    & 21\,$\pm$\,2    &8.9\,$\pm$\,0.3     &0.90 \\  
21       & 126\,$\pm$\,7   &3.2\,$\pm$\,0.4  & 7.9\,$\pm$\,0.5   &27\,$\pm$\,2    &290\,$\pm$\,20 &83\,$\pm$\,5   & 29\,$\pm$\,2    & 20.2\,$\pm$\,1.4& 8.5\,$\pm$\,0.3    &0.98 \\  
22$^{*}$ & 220\,$\pm$\,20 &23\,$\pm$\,2     & 62\,$\pm$\,3      &38\,$\pm$\,2    &290\,$\pm$\,20 &123\,$\pm$\,8  & 64\,$\pm$\,4    & 46\,$\pm$\,3    &2.03\,$\pm$\,0.07   &0.61 \\ 
23       & 130\,$\pm$\,8  &6.6\,$\pm$\,0.6  & 18.0\,$\pm$\,1.1  &25\,$\pm$\,2    &290\,$\pm$\,20 &73\,$\pm$\,5   & 31\,$\pm$\,2    & 21\,$\pm$\,2    &8.9\,$\pm$\,0.3     &0.90 \\  
24       & 300\,$\pm$\,20  & 22.1\,$\pm$\,1.2& 64\,$\pm$\,3      &31\,$\pm$\,2    &290\,$\pm$\,20 &91\,$\pm$\,6   & 48\,$\pm$\,3    & 33\,$\pm$\,2    &2.80\,$\pm$\,0.09   &0.42 \\  
27       & 210\,$\pm$\,20  &...          & ...           &19\,$\pm$\,2    &280\,$\pm$\,30 &68\,$\pm$\,6   & 42\,$\pm$\,4    & 27\,$\pm$\,3    &3.8\,$\pm$\,0.1     &0.79 \\  
28       &   ...       & 74.6\,$\pm$\,3.6& 220\,$\pm$\,10    &13.3\,$\pm$\,0.9&290\,$\pm$\,20 &43\,$\pm$\,3   & 15\,$\pm$\,1    &10.2\,$\pm$\,0.7 &14.4\,$\pm$\,0.05    &1.03 \\ 
30       & ...         & ...         & 7.3\,$\pm$\,0.3   &20.6\,$\pm$\,1.3&290\,$\pm$\,20 &62\,$\pm$\,4   & 25\,$\pm$\,2    & 16.9\,$\pm$\,1.1&2.52\,$\pm$\,0.08   &0.19 \\   
31       & 290\,$\pm$\,20  & 11.2\,$\pm$\,0.8& 35\,$\pm$\,2      &28\,$\pm$\,2    &290\,$\pm$\,20 &83\,$\pm$\,5   & 38\,$\pm$\,3    & 28\,$\pm$\,2    &5.9\,$\pm$\,0.2     &0.57 \\ 
36       & 253\,$\pm$\,19  &6.3\,$\pm$\,0.3  & 15.6\,$\pm$\,0.7  &21\,$\pm$\,2    &280\,$\pm$\,30 &56.5\,$\pm$\,5 &...          &...          &7.7\,$\pm$\,0.3     &1.03 \\  
37       & 257\,$\pm$\,13  & 16.2\,$\pm$\,0.8&  48\,$\pm$\,2     &25\,$\pm$\,2    &280\,$\pm$\,20 &75\,$\pm$\,4   &...          &...          &13.3\,$\pm$\,0.5    &0.74 \\ 
\hline	 
\end{tabular}
{\footnotesize \begin{minipage}{\textwidth}
\begin{tabular}{l}
$^{*}$  SNRs candidates (see Sect. \ref{SNRs}) for details. \\
 \end{tabular}
\end{minipage}}
\end{minipage}
\end{table*}
%%%%%%%%%%%%%%%%%%%%%%%%%%%%%%%%%%%%%%%%%%%%%%%%%%%%%%%%%%%%%%%%%%%%%%%%%%%%%%%%%%%%%%%%%%%%%%%%%%%%%%%%%%%%%%%%%%%%%%%%%%%%%%%%%%%%%%%%%%%%%%%%%%%%%%%%%%%%%%%%%%%%%%%%%%%%%%%%%%%%%%%%%%%%%%%%%%%%%%%%%%%%%%%

\section{Physical conditions of the ionized gas}
\label{Te}
\begin{figure}
   \centering
   \includegraphics[bb= 127 180 460 594, clip, width=7.3cm]{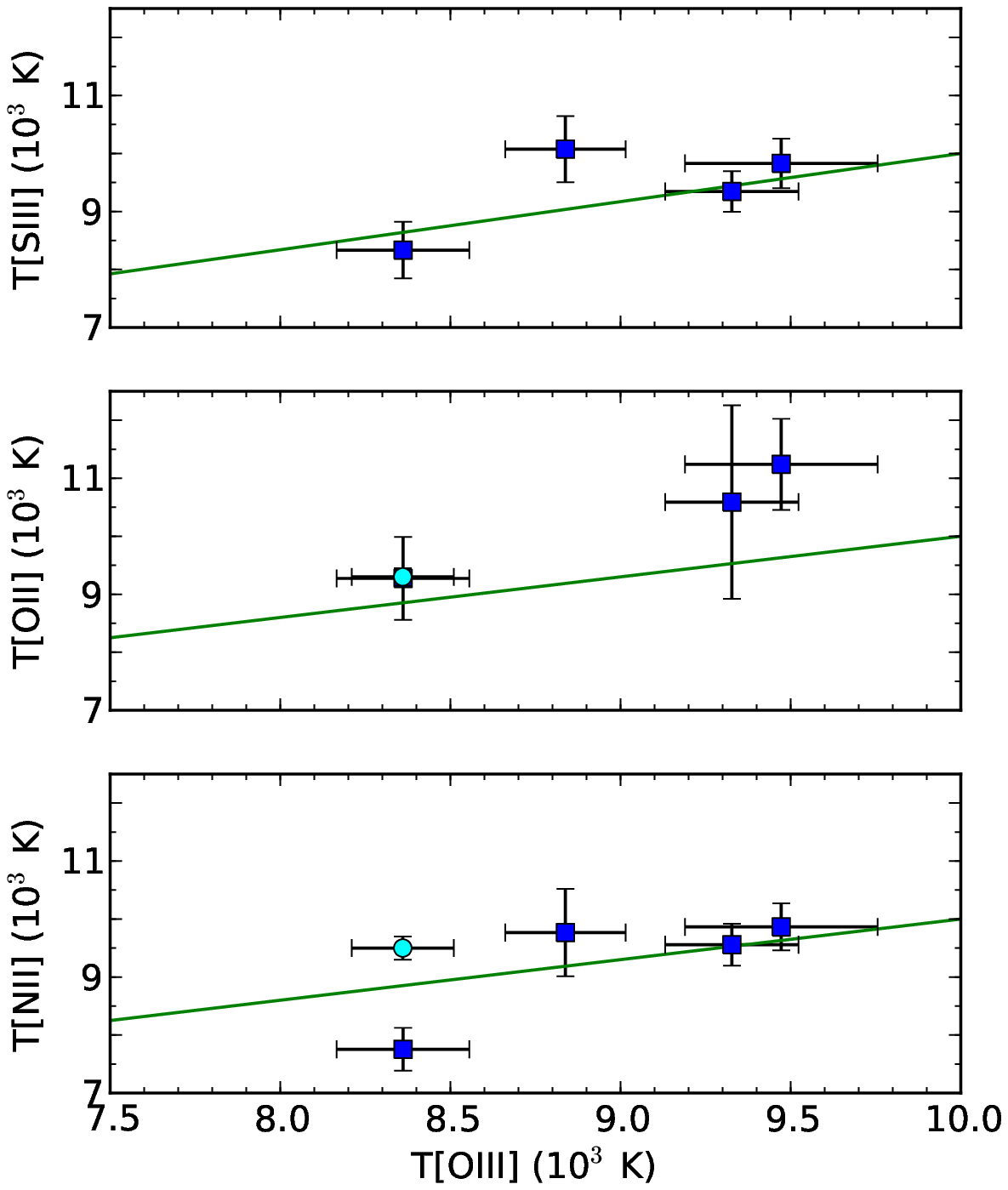}
   \caption{Comparison between the electron temperature obtained from the [\oiii]$\lambda$4363/[\oiii]$\lambda\lambda$4959,5007 
line ratio  and the electron temperatures obtained from [\siii]$\lambda$6312/[\siii]$\lambda\lambda$9069,9532 (top), 
[\oii]$\lambda$7325/[\oii]$\lambda\lambda$3727,3729 (middle), and [\nii]$\lambda$5755/[\nii]$\lambda\lambda$6548,6584 (bottom).  The cyan
circle corresponds to the \hii\ region K932, studied by \citet{esteban09}.
The green straight lines indicate the relationship between different temperatures predicted by the  \citet{garnett92} models.}
\label{Te_a}
\end{figure}
\begin{figure}
   \centering
   \includegraphics[bb= 127 180 460 594, clip,width=7.3cm]{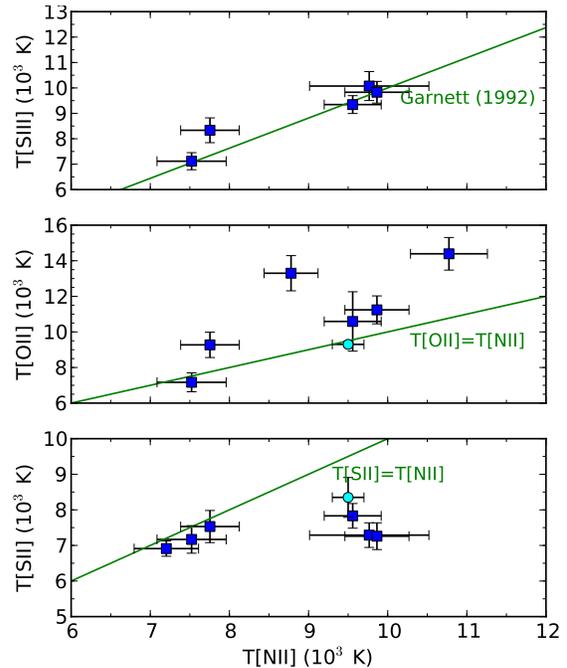}
   \caption{Comparison between the electron temperature obtained from the [\nii]$\lambda$5755/[\nii]$\lambda\lambda$6548,6584 
line ratio  and the electron temperatures obtained from [\siii]$\lambda$6312/[\siii]$\lambda\lambda$9069,9532 
(top), [\oii]$\lambda$7325/[\oii]$\lambda\lambda$3727,3729 (middle) and [\sii]$\lambda$4072/[\sii]$\lambda\lambda$6717,6731 
(bottom). The cyan
circle corresponds to the \hii\ region K932, studied by \citet{esteban09}.}   
\label{Te_b}
\end{figure}
The physical conditions of the ionized gas (electron temperature and density) 
have been determined from collisionally excited line ratios.
The {\tt temden} IRAF task within the {\tt NEBULAR} package \citep{nebular}
was employed. The  {\tt temden} task is an extension of the FIVEL program 
\citep*{DeRobertis87} for a larger set of ions, emission lines and  
number of levels (between 5 and 8 depending on the ion).
The atomic parameters (transition probabilities and collisional strengths) 
for the ions of interest were updated with more recent data as shown in Table~5 
of \citet{Bresolin09}.

The derivation of the electron temperature of an \hii\ region requires the detection of auroral lines 
which arise from the transition from the second  lowest excited level to the 
lowest excited level.
Auroral lines are generally faint and difficult to detect, with strongly decreasing intensities with increasing gas metallicity. This explains why,
to our knowledge, there is surprisingly only one previous published measurement of 
electron temperature for an \hii\ region in M31: region K932, analysed by \citet{esteban09} from Keck HIRES data.
For nine \hii\ regions of our sample, one or more of the auroral lines 
[\oiii]$\lambda$4363, [\siii]$\lambda$6312, [\oii]$\lambda$7325, [\nii]$\lambda$5755 and
[\sii]$\lambda$4072 have been detected.
The ratio of these lines to stronger nebular lines of the same ions are  highly dependent 
on electron temperature and can thus be employed for measuring it.
The auroral line [\nii]$\lambda$5755 was detected in all nine \hii\ regions, while [\siii]$\lambda$6312
was measured in only five. This is in part due to the fact that this emission line, due 
to the radial velocity of M31,  is within 2\,\AA\ of the telluric line [\oi]$\lambda$6300\,\AA, and 
its firm detection is therefore subject to higher uncertainties.

Density and temperature of the ionized gas have been simultaneously obtained from an iterative process.
The temperature was initially set to $T_e$=7500~K and a preliminary value for the electron density 
$N_e$ was derived from the 
[\sii]$\lambda$6717/[\sii]$\lambda$6731 line ratio.
This initial value of N$_e$ was then used for computing $T_e$ for the different 
auroral--to--nebular line ratios. The process was repeated until convergence. An initial estimate
of the ionic abundances was then obtained (as described in Sect.~\ref{ionic}) and from these we 
estimated the contamination of the auroral lines [\nii]$\lambda$5755 and [\oii]$\lambda$7325 due 
to recombination following \citet{liu2000} (their Eqs. 1 and 2) and assuming 
N$^{++}$/N$^{+}$~$\simeq$~O$^{++}$/O$^{+}$ to obtain N$^{++}$/H$^{+}$.
The electron temperatures and densities were  recomputed until convergence after the correction
for recombination contamination was applied. This correction is small and varies between 0.1 
and 0.9\% for [\nii]$\lambda$5755/\hb\  and between 0.3 and 3.7\% for [\oii]$\lambda$7325/\hb.
We note that in a number of regions $T$[\oii] is above the range of validity of the \citet{liu2000}
formula for [\oii]$\lambda$7325/\hb\ (5000~K $<$ $T_e$ $<$ 10000~K). %and therefore the correction applied could not be appropriated. 
In any case, the errors in the derived $T$[\oii] value for these 
\hii\ regions are significantly larger than the corresponding correction.
\begin{table*}
\begin{minipage}{120mm}
\centering
\caption{Measured \hii\ region electron temperatures (in K) and densities (in cm$^{-3}$). The auroral line intensities [\oii]$\lambda$7325 and  [\nii]$\lambda$5755 have been corrected for recombination contamination following \citet{liu2000}.}
\label{MeasuredTe}
\begin{tabular}{l c c c c c c }     % 7 columns
\hline
ID & $T$[\oiii]                 &  $T$[\siii]$^{*}$                 & $T$[\oii]            &  $T$[\nii]                  & $T$[\sii]                          &  $N_e$            \\[1mm]
              & $\frac{4363}{4959+5007}$  & $\frac{6312}{9069+9532}$  & $\frac{7325}{3727}$ &  $\frac{5755}{6548+6583}$ & $\frac{4072}{6717+6731}$ &      [\sii]$\lambda\lambda$6717/6731      \\[1.5mm]
%      &     &          &        &           &       &       \\
\hline
 2 &    ...        &7100\,$\pm$\,300  & 7200\,$\pm$\,500    &  7500\,$\pm$\,400  &  7200\,$\pm$\,400 &66         \\
 3 &    ...        &    ...       &     ...         &  8300\,$\pm$\,300  &      ...      &49         \\
17 &  8400\,$\pm$\,200 & 8300\,$\pm$\,500 & 9300\,$\pm$\,700    &  7800\,$\pm$\,400  & 7500\,$\pm$\,500 &38         \\ 
19 &   ...         &    ...       &     ...         &  7200\,$\pm$\,400  & 6900\,$\pm$\,200 &24$^{**}$   \\
25 &  9300\,$\pm$\,200 & 9300\,$\pm$\,400 & 10600\,$\pm$\,1700 &  9600\,$\pm$\,400  &  7800\,$\pm$\,300 &28         \\
26 & 9500\,$\pm$\,300  & 9800\,$\pm$\,400 &  11200\,$\pm$\,800 &  9900\,$\pm$\,400  &  7300\,$\pm$\,400 &17$^{**}$   \\
33 &    ...        &     ...      & 13300\,$\pm$\,1000  &  8800\,$\pm$\,300  &    ...        &24          \\ 
34 &     ...       &     ...      & 14400\,$\pm$\,900   & 10800\,$\pm$\,500  &    ...        &14$^{**}$   \\  
35 &  8800\,$\pm$\,200 &10100\,$\pm$\,600 &    ...          &  9800\,$\pm$\,800  &  7300\,$\pm$\,300 &80          \\
\hline
\end{tabular}
 {\footnotesize \begin{tabular}{l}
$^{*}$  Assuming [\siii]$\lambda$9532/[\siii]$\lambda$9069=2.44.\\
$^{**}$ [\sii]6717/[\sii]6731 line ratios slightly above 1.41 (1.44, 1.43 and 1.45 respectively); it\\
 has been assumed that [\sii]$\lambda$6717/[\sii]$\lambda$6731=1.41.\\
\end{tabular}}
\end{minipage}
\end{table*}

The electron temperatures are shown in columns 2 to 6 of Table~\ref{MeasuredTe}.
The errors reported for  $T_e$ arise from error propagation of the relevant line flux errors and 
are in the range 200--600~K for all ions, except in the case of $T$[\oii], for which the errors 
are larger (400--1700~K). This is mainly due to the observational uncertainty 
that involves the measurement of the [\oii] multiplet at 7325\,\AA, located in a spectral 
region of strong OH airglow emission. 

Figs.~\ref{Te_a} and~\ref{Te_b} show a comparison between 
the different ionic temperatures we measured and the results one would obtain from
model predictions by  \citet{garnett92}, frequently used for predicting ionic 
temperatures when only one or two auroral lines are detected.
Fig.~\ref{Te_a} shows the relations between $T$[\siii], $T$[\oii] and $T$[\nii] with 
respect to $T$[\oiii]. For most of the regions with a $T$[\oiii] determination, the 
\citet{garnett92}  temperature relations %T[\siii]--T[\oiii] and T[\oii]--T[\oiii]:
\begin{equation}
\label{eq1}
\centering
T\mbox{[\siii]}=0.83 T\mbox{[\oiii]} + 1700 \mbox{K}
\end{equation}
\begin{equation}
\label{eq2}
\centering
T\mbox{[\oii]}=T\mbox{[\nii]}=0.70 T\mbox{[\oiii]} + 3000 \mbox{K} 
\end{equation}
appear to be satisfactorily followed by the empirical data, especially for the case of $T$[\siii] vs. $T$[\oiii]. In the other two cases
the data points deviate more significantly from the theoretical relations.
This was also found by other authors with a larger sample of \hii\ regions \citep*[e.g.][]{kennicutt03, montero-diaz03}.

Fig.~\ref{Te_b} shows the relations between the different ionic temperatures with respect
to $T$[\nii], and highlights a result already found by other authors \citep[e.g.][]{bresolin05}: 
$T$[\oii] seems to overestimate the temperature with respect to $T$[\nii], while the opposite happens 
to $T$[\sii], which gives always lower temperatures than $T$[\nii].
However, $T$[\siii] seems to be well correlated with $T$[\nii], following closely the model predictions by \citet{garnett92}.

The  electron densities $N_e$ are shown in column~7 of Table~\ref{MeasuredTe}. All 
\hii\ regions are in the low--density regime, with densities  below 100~cm$^{-3}$.
For three \hii\ regions we measured a line ratio [\sii]$\lambda$6717/[\sii]$\lambda$6731
slightly above the theoretical limit. For those regions we assumed 
a line ratio equal to 1.41 (equivalent to the theoretical limit).

\section{Chemical abundances}
\label{abundances}
\subsection{Ionic abundances}
\label{ionic}
The calculation of ionic abundances from collisionally--excited lines requires 
a good knowledge of the electron temperature. An onion--model  is commonly 
adopted to describe the ionization structure of an \hii\ region, with a 
number of zones of different $T_e$ where atomic species of similar excitation coexist. 
In light of the correlations found in Sec.~\ref{Te} between the $T_e$ obtained from different atomic species, we 
have adopted a two--zone representation for describing the temperature stratification of the \hii\ regions:
a low--excitation zone, characterized by $T$[\nii], where ions of low ionization potential coexist
(O$^+$, S$^+$, N$^+$) and a high--excitation zone for the ions with higher ionization potential 
(O$^{++}$, S$^{++}$, Ar$^{++}$, Ne$^{++}$), characterized by $T$[\oiii]. 
For the \hii\ regions for which we could not measure $T$[\oiii], we have used the average of our measured value for
$T$[\siii], and the value for $T$[\oiii] obtained from the measured $T$[\nii] by inverting Eq.~\ref{eq2} \citep{garnett92}.
When $T$[\siii] was not available, $T$[\oiii] has been  obtained from $T$[\nii] and the mentioned relation (Eq.~\ref{eq2}).
Table~\ref{AdoptedTe} contains the  temperatures adopted for the ionic abundance calculation for each \hii\ region.
The errors quoted in the table correspond to the errors in the measured values for $T$[\nii] and $T$[\oiii], as described 
in Sect.~\ref{Te}, when these temperatures were available. When the Garnett relations were used, the uncertainty in $T$[\oiii]
comes from error propagation of the uncertainty in $T$[\nii], $T$[\siii] and an additional term added in quadrature, which is an 
estimate of the uncertainty in the scaling relation ($\pm$400K for
$T$[\nii] vs. $T$[\oiii]) obtained from works with a large number of  \hii\ regions \citep{Bresolin09}.

\begin{table}
\centering
\caption{Adopted electron temperatures for the two--zone scheme of the \hii\ region ionization structure.}
\label{AdoptedTe}
\begin{tabular}{l c c  }     % 3 columns
\hline
ID & $T$[O$^+$, N$^+$, S$^+$]  & $T$[S$^{++}$, Ar$^{++}$, O$^{++}$, Ne$^{++}$]  \\
   &      (K)             &     (K)                    \\
\hline
 2 &   7500\,$\pm$\,400 & 6800\,$\pm$\,400 \\
 3 &   8300\,$\pm$\,300 & 7600\,$\pm$\,600  \\
17 &   7800\,$\pm$\,400 & 8400\,$\pm$\,200 \\
19 &   7200\,$\pm$\,400 & 6000\,$\pm$\,700 \\
25 &   9600\,$\pm$\,400 & 9300\,$\pm$\,200 \\
26 &   9900\,$\pm$\,400 & 9500\,$\pm$\,300 \\
33 &   8800\,$\pm$\,300 & 8300\,$\pm$\,600  \\
34 &  10800\,$\pm$\,500 &11100\,$\pm$\,800 \\ 
35 &   9800\,$\pm$\,800 & 8800\,$\pm$\,200 \\
\hline
\end{tabular}
{\footnotesize \begin{tabular}{l}
\end{tabular}}
\end{table}

The {\tt ionic} IRAF task within the {\tt NEBULAR} package \citep{nebular}
was employed for obtaining the ionic abundances of O$^+$, O$^{++}$, N$^+$, S$^+$, 
S$^{++}$, Ar$^{++}$ and Ne$^{++}$. They were estimated from the relevant reddening-corrected  
emission line ratios from those species (Table~\ref{lines}) and the corresponding
electron temperatures of the adopted two--zone scheme (Table~\ref{AdoptedTe}).
Table~\ref{ionic_abundances} shows the resulting ionic abundances with respect to H$^+$.

\begin{table*}
\centering
\begin{minipage}{140mm}
\caption{O$^+$/O  and ionic abundances (columns 3 to 9) given as  $12+\log(\mbox{X}/\mbox{H})$, where X is the ion in the table header.}
\label{ionic_abundances}
\centering
\begin{tabular}{r c c c c c c c c }    
\hline	
ID & O$^+$/O  &  O$^+$ &  O$^{++}$  & N$^+$  & S$^+$  & S$^{++}$ &  Ne$^{++}$ & Ar$^{++}$ \\
\hline    
2  & 0.8\,$\pm$\,0.3 &8.50\,$\pm$\,0.14 &8.02\,$\pm$\,0.13 &7.56\,$\pm$\,0.09 &6.30\,$\pm$\,0.08 &6.95\,$\pm$\,0.07 &7.21\,$\pm$\,0.16 &6.22\,$\pm$\,0.09 \\
3  & 0.8\,$\pm$\,0.2 &8.16\,$\pm$\,0.07 &7.66\,$\pm$\,0.14 &7.41\,$\pm$\,0.04 &6.16\,$\pm$\,0.04 &6.78\,$\pm$\,0.08 &    ...       &6.06\,$\pm$\,0.10 \\
17 & 0.7\,$\pm$\,0.2 &8.25\,$\pm$\,0.11 &7.92\,$\pm$\,0.04 &7.40\,$\pm$\,0.07 &6.07\,$\pm$\,0.06 &6.73\,$\pm$\,0.02 &7.03\,$\pm$\,0.05 &6.11\,$\pm$\,0.03 \\
19 & 0.7\,$\pm$\,0.3 &8.49\,$\pm$\,0.15 &8.06\,$\pm$\,0.31 &7.70\,$\pm$\,0.09 &6.46\,$\pm$\,0.08 &7.03\,$\pm$\,0.16 &    ...       &6.36\,$\pm$\,0.21 \\
25 & 0.4\,$\pm$\,0.1 &7.86\,$\pm$\,0.07 &8.08\,$\pm$\,0.03 &6.83\,$\pm$\,0.05 &5.68\,$\pm$\,0.04 &6.59\,$\pm$\,0.02 &7.34\,$\pm$\,0.04 &6.03\,$\pm$\,0.02 \\
26 & 0.5\,$\pm$\,0.1 &7.95\,$\pm$\,0.08 &8.00\,$\pm$\,0.05 &6.98\,$\pm$\,0.05 &5.85\,$\pm$\,0.04 &6.58\,$\pm$\,0.03 &7.26\,$\pm$\,0.06 &6.01\,$\pm$\,0.03 \\
33 & 0.8\,$\pm$\,0.2 &8.17\,$\pm$\,0.08 &7.65\,$\pm$\,0.14 &7.29\,$\pm$\,0.05 &6.18\,$\pm$\,0.05 &6.67\,$\pm$\,0.08 &   ...        &5.91\,$\pm$\,0.09 \\
34 & 0.6\,$\pm$\,0.1 &7.90\,$\pm$\,0.08 &7.68\,$\pm$\,0.10 &7.19\,$\pm$\,0.05 &6.13\,$\pm$\,0.05 &6.24\,$\pm$\,0.06 &6.63\,$\pm$\,0.12 &5.93\,$\pm$\,0.07 \\
35 & 0.3\,$\pm$\,0.1 &7.80\,$\pm$\,0.16 &8.27\,$\pm$\,0.03 &6.84\,$\pm$\,0.09 &5.72\,$\pm$\,0.09 &6.64\,$\pm$\,0.02 &7.55\,$\pm$\,0.04 &6.13\,$\pm$\,0.02 \\
\hline	
\end{tabular}
\end{minipage}
\end{table*}

\subsection{Total abundances}
\label{total}

\begin{table*}
\begin{minipage}{126mm}
\caption{Total abundances.}
\label{total_abundances}
\begin{tabular}{r c c c c c c c }   
\hline	
 ID & 12+$\log$(O/H) & $\log{\mbox{(N/O)}}$ & $\log{\mbox{(S/O)}}$&$\log{\mbox{(Ar/O)}}$& $\log{\mbox{(Ne/O)}}$&  He/H  &R/R$_{25}$$^{*}$\\
\hline
 2 &   8.62\,$\pm$\,0.11  &   $-$0.95\,$\pm$\,0.17   &  $-$1.58\,$\pm$\,0.13   &  $-$2.32\,$\pm$\,0.14  &   $-$0.81\,$\pm$\,0.20   &0.098\,$\pm$\,0.027  &0.40\\
 3 &   8.28\,$\pm$\,0.06  &   $-$0.76\,$\pm$\,0.08   &  $-$1.41\,$\pm$\,0.09   &  $-$2.14\,$\pm$\,0.12  &        ...         &0.111\,$\pm$\,0.024  &0.42\\
17 &   8.42\,$\pm$\,0.08  &   $-$0.86\,$\pm$\,0.13   &  $-$1.59\,$\pm$\,0.08   &  $-$2.23\,$\pm$\,0.08  &   $-$0.89\,$\pm$\,0.06   &0.111\,$\pm$\,0.017  &0.44\\
19 &   8.63\,$\pm$\,0.14  &   $-$0.79\,$\pm$\,0.17   &  $-$1.49\,$\pm$\,0.19   &  $-$2.19\,$\pm$\,0.25  &        ...         &0.083\,$\pm$\,0.034  &0.46\\
25 &   8.28\,$\pm$\,0.04  &   $-$1.02\,$\pm$\,0.09   &  $-$1.58\,$\pm$\,0.04   &  $-$2.23\,$\pm$\,0.04  &   $-$0.74\,$\pm$\,0.05   &0.088\,$\pm$\,0.005  &0.71\\
26 &   8.28\,$\pm$\,0.04  &   $-$0.98\,$\pm$\,0.09   &  $-$1.59\,$\pm$\,0.05   &  $-$2.23\,$\pm$\,0.06  &   $-$0.73\,$\pm$\,0.07   &0.090\,$\pm$\,0.006  &0.71\\
33 &   8.29\,$\pm$\,0.07  &   $-$0.89\,$\pm$\,0.10   &  $-$1.49\,$\pm$\,0.09   &  $-$2.30\,$\pm$\,0.12  &        ...         &0.074\,$\pm$\,0.015  &0.76\\
34 &   8.10\,$\pm$\,0.06  &   $-$0.71\,$\pm$\,0.10   &  $-$1.60\,$\pm$\,0.07   &  $-$2.14\,$\pm$\,0.09  &   $-$1.05\,$\pm$\,0.16   &0.089\,$\pm$\,0.004  &0.75\\
35 &   8.40\,$\pm$\,0.05  &   $-$0.96\,$\pm$\,0.18   &  $-$1.60\,$\pm$\,0.05   &  $-$2.24\,$\pm$\,0.05  &   $-$0.72\,$\pm$\,0.05   &0.087\,$\pm$\,0.005  &0.76\\
\hline	
\end{tabular}
\medskip
{\footnotesize \begin{tabular}{l}
$^{*}$ Deprojected galactocentric distance normalized to R$_{25}$. See Table~\ref{M31 params} for the M31 adopted parameters.\\
\end{tabular}}
\end{minipage}
\end{table*}
The total abundance relative to hydrogen of a given element is given by the sum of the abundances of all its ions.
However, not all ions have emission lines in our spectra and we have to correct for these
unseen ionization stages using ionization correction factors (ICFs).
We have used the following common assumptions:\\[-2mm]

{\bf Oxygen.} \heii~$\lambda$4686 is only detected in one target (\#8) which is a SNR. 
This  implies  that all \hii\ regions are of sufficiently low excitation that the amount of oxygen in  
ionization stages higher than  O$^{++}$ is negligible.
We have therefore set O/H = O$^{+}$/H$^{+}$ + O$^{++}$/H$^{+}$.\\[-2mm]

{\bf Nitrogen and Neon.} For these elements we have assumed that N/O\,=\,N$^{+}$/O$^{+}$ and
Ne/O\,=\,Ne$^{++}$/O$^{++}$, which result from ionization potential considerations of the ions involved \citep[see][]{Bresolin09}.\\[-2mm]
%% Buscar mas sobre esto 
\begin{figure}
   \centering
   \includegraphics[width=8.5cm]{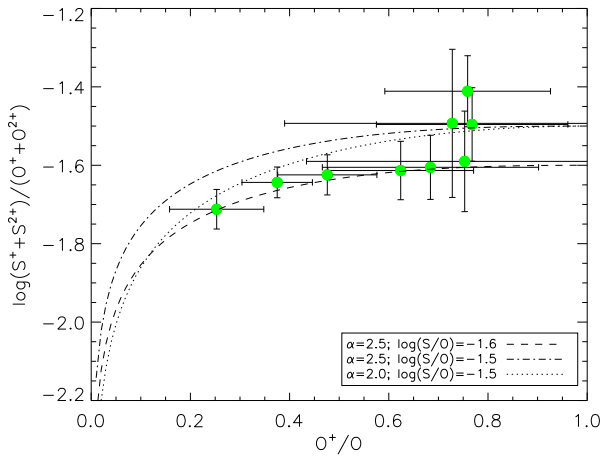}
   \caption{Logarithm of (S$^{+}$\,+\,S$^{++}$)/(O$^{+}$\,+\,O$^{++}$) versus O$^{+}$/O for our 
    \hii\ region sample (filled circles). The lines show the sulfur ionization correction factor (Eq.~\ref{Sfactor})
    for different values of $\alpha$ and assuming a constant value for $\log$(S/O).}
\label{ICF_S}
\end{figure}

{\bf Argon.} Only Ar$^{++}$ is observed in our optical spectra. However,
the amount of Ar$^{+}$ in low excitation  \hii\ regions (our case) can be important \citep*{b04}.
We have employed the ICFs obtained from photoionization models by Izotov et al. (2006).
%\citet{Izotov06}. 
These authors proposed different parameterizations depending on metallicity. Those for high ($12+\log$(O/H) $ > 8.2$) and 
intermediate ($7.6 < 12+\log$(O/H) $\le 8.2$) metallicity cover our range of oxygen abundances:
\begin{equation}
\mbox{ICF(Ar$^{++}$)} = \left\{ \begin{array}{rl}
 0.285 v +0.833 + 0.051/v &\mbox{    intermediate Z} \\
 0.517 v +0.763 + 0.042/v &\mbox{    high Z} 
\end{array} \right.
\end{equation}
where $v\equiv\mbox{O}^{+}$/(O$^{+}$+O$^{++}$). \\[-2mm]

{\bf Sulfur.} In the case of low excitation \hii\ regions the ICF to correct 
(S$^{+}$\,+\,S$^{++}$)/H$^{+}$ for the presence of higher ionization sulfur stages is expected
to be small. We have used the ICF given by the formula \citep{peimbert69,stasinska78,french81}:
\begin{equation}
\label{Sfactor}
\mbox{ICF(S$^{+}$\,+\,S$^{++}$)} = \frac{\mbox{S}}{\mbox{S$^{+}$\,+\,S$^{++}$}} = \biggl[1-\biggl(1-\frac{\mbox{O$^{+}$}}{\mbox{O}}\biggl)^\alpha\biggl]^{-1/\alpha}
\end{equation}
where we have employed $\alpha=2.5$ \citep{b04}. Fig.~\ref{ICF_S} shows $\log$[(S$^{+}$\,+\,S$^{++}$)/(O$^{+}$\,+\,O$^{++}$)]
for our \hii\ region sample versus the oxygen fractional ionization O$^{+}$/O. Eq.~\ref{Sfactor} has been overplotted 
for $\alpha=2$ and $\alpha=2.5$ and assuming $\log$(S/O)\,=\,$-1.5$ and $\log$(S/O)\,=\,$-1.6$  (representative values for our sample). 
Apart from three low-excitation regions with considerable error bars, the curve with $\alpha=2.5$ and 
$\log$(S/O)\,=\,$-1.6$ follows the data rather well.
We have then kept this parameter choice, in agreement with previous work on extragalactic \hii\ regions \citep{kennicutt03,b04}.
This equation yields corrections of 16-30\% for the highest excitation \hii\ regions of the sample (O$^{+}$/O\,$\sim$\,0.25-0.4),
2-9\% for \hii\ regions with O$^{+}$/O\,$\sim$\,0.5-0.7, and corrections around 1\% for the lowest excitation regions with 
O$^{+}$/O\,$\sim$\,0.8.

%Esto esta calculado con diagramas_diagnostico_v1.pro

The total abundances of O, N, S, Ar and Ne, obtained as described above, are summarized in Table~\ref{total_abundances}.

\subsection{Abundance trends with log(O/H) and  galactocentric radius}
\begin{figure}
   \centering
   \includegraphics[bb=20 190 326 600, clip, width=8.2cm]{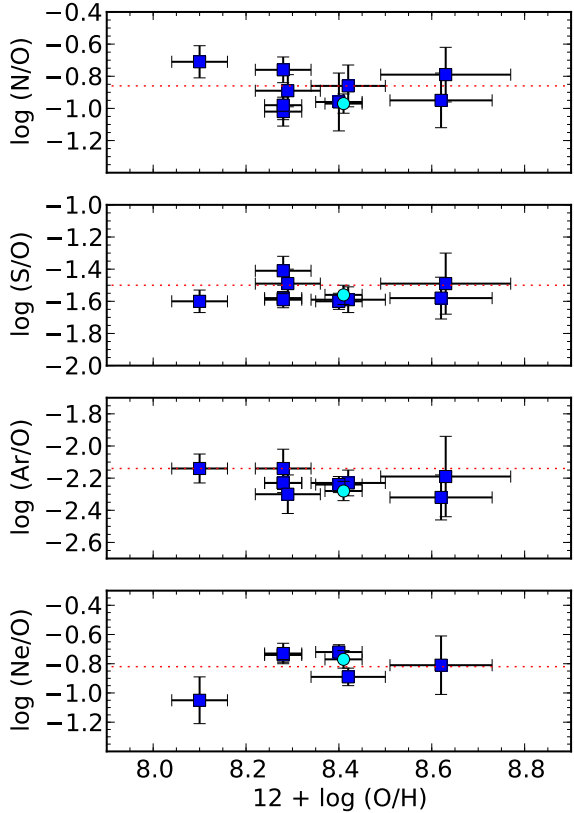}
   \caption{Trends in the abundance ratios relative to oxygen  with 12+log(O/H). The horizontal dotted lines indicate the 
corresponding solar value according to  \citet{Lodders03}. The filled circle corresponds to the \hii\ region K932 with 
abundances measured by \citet{esteban09}.}
\label{total_abund_O}
\end{figure}

\begin{figure}
   \centering
   \includegraphics[bb= 20 190 326 600, clip, width=8.2cm]{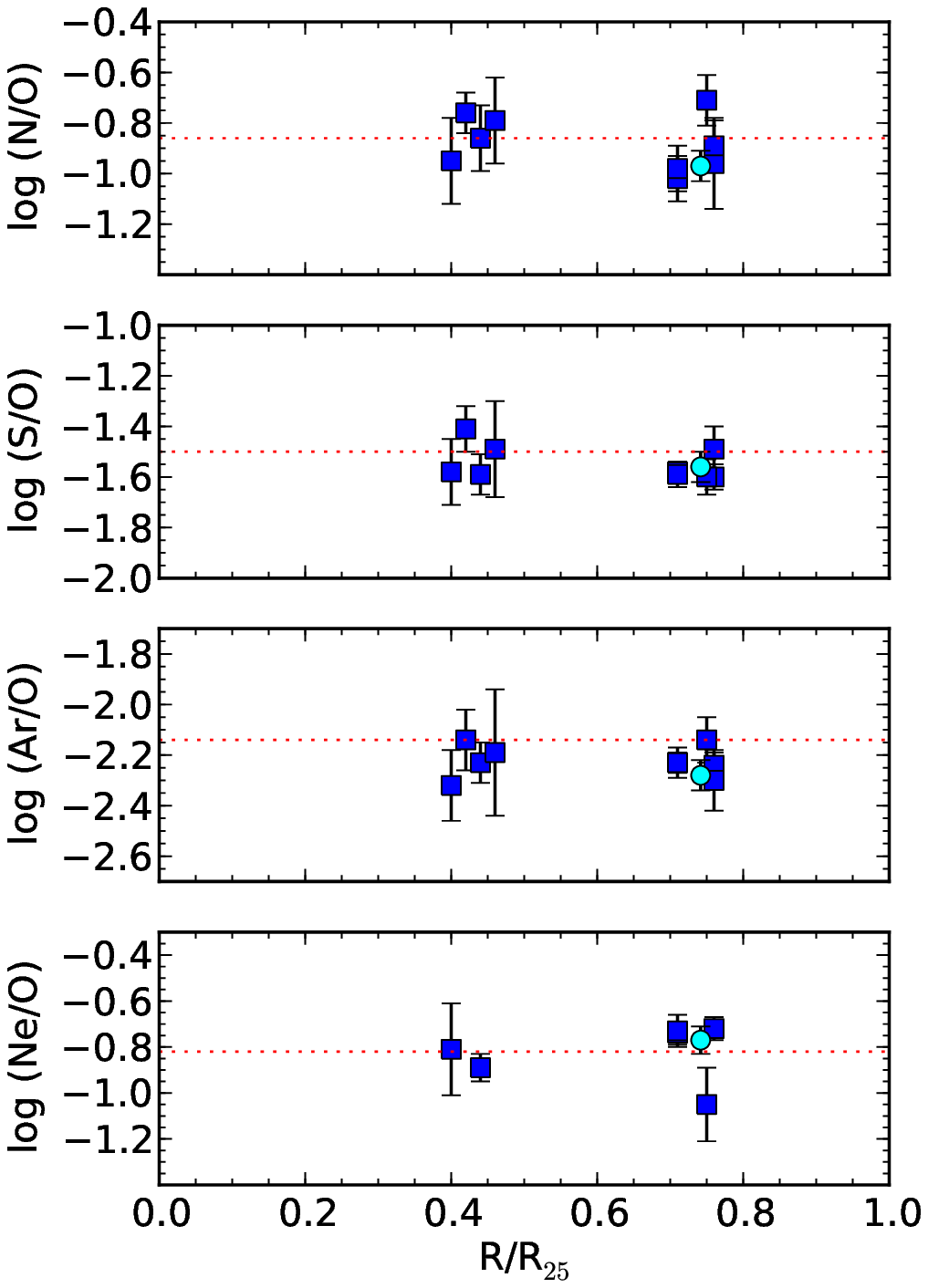}
   \caption{Trends in the abundance ratios relative to oxygen with galactocentric radius. The horizontal dotted lines 
indicate the corresponding solar value according to  \citet{Lodders03}. The filled circle corresponds to the \hii\ 
region K932 with abundances measured by \citet{esteban09}.}   
\label{total_abund_R}
\end{figure}

Fig.~\ref{total_abund_O} shows the total abundances of nitrogen, sulfur, neon and argon with respect to oxygen
as a function of the oxygen abundances.
The N/O ratio does not show the typical increase with 12+log(O/H) expected for a pure secondary origin of 
nitrogen \citep{vila-costas93}. The plot has a considerable scatter ($\sim0.11$~dex) around a mean corresponding to the solar 
value, log(N/O)$_\odot=-0.86$ \citep{Lodders03}, and seems more compatible with a primary-plus-secondary origin 
for this element. The  dispersion  found in this relation is often explained as due to a time delay between 
the release of nitrogen and the release of oxygen back into the interstellar medium by the respective stellar 
sources \citep{vila-costas93, mallery07}. 
Also \hii\ regions in the galaxy M33 display an approximately constant value of the N/O ratio \citep{B10}, 
with a mean log(N/O)$\sim-1.2$, i.e.~between the solar value  and typical values
for dwarf galaxies.
The flat radial distribution of the N/O ratio is shown in the top panel of Fig.~\ref{total_abund_R}.

The S/O ratio is  constant as a function of 12+log(O/H), with a mean value log(S/O)\,=\,$-1.55\pm0.07$, consistent, within errors, 
with the solar value  \citep[$-1.50\pm0.06$, ][]{Lodders03}. 
The Ar/O ratio is also essentially constant with oxygen abundance, but with a mean value of log(Ar/O)=$-2.22\pm0.06$, 
slightly subsolar but marginally compatible with the solar value \citep[$-2.14\pm0.09$,][]{Lodders03} within the uncertainties.
The Ne/O ratio, with a mean value of log(Ne/O)\,=\,$-$0.82 and a standard deviation of $\sim0.13$~dex also shows 
a rather flat distribution in the whole range of oxygen abundance measured in our sample of \hii\ regions, 
and matches the solar value \citep[$-0.82\pm0.11$,][]{Lodders03}.

Our data are compatible with a flat distribution of the S/O, Ar/O and Ne/O ratios with galactocentric distance 
(see Figure~\ref{total_abund_R}), but since our \hii\ region data are concentrated at only
two different galactocentric distances further data would be required to confirm these trends.

\subsection{He abundance}

Our \hii\ region spectra contain a number of \hei\ recombination emission lines
that can be used to estimate the He$^+$ abundance, as the emission line is proportional to the
line emissivity and to the corresponding ionic abundance.
\heii~$\lambda4686$  is not detected in any \hii\ region of the sample and therefore the 
contribution of He$^{++}$ to the total He abundance can be neglected.
\begin{table*}
\begin{minipage}{126mm}
\caption{He$^+$ abundance.}
\label{abund_he}
\centering
\begin{tabular}{@{}l c c c c  c c @{}}     % 7 columns
\hline	
ID &    \multicolumn{6}{c}{He$^+$/H$^+$}  \\ 

           &   $\lambda$4026    &   $\lambda$4471     &  $\lambda$4922     &   $\lambda$5876    &  $\lambda$6678      &   Average\footnote{Weighted average of the He$^+$/H$^+$ abundance ratio computed from all  \hei\ lines in the table ($\lambda$4026, $\lambda$4471, $\lambda$4922, $\lambda$5876, $\lambda$6678).}\\
\hline
2  &  0.056\,$\pm$\,0.005  &    0.063\,$\pm$\,0.006  &   0.066\,$\pm$\,0.006  &   0.069\,$\pm$\,0.010  &   0.071\,$\pm$\,0.009   &   0.066\,$\pm$\,0.004\\
3  &       ...         &    0.074\,$\pm$\,0.006  &   0.076\,$\pm$\,0.006  &   0.075\,$\pm$\,0.007  &   0.079\,$\pm$\,0.007   &   0.076\,$\pm$\,0.003\\  
17 &  0.086\,$\pm$\,0.010  &    0.081\,$\pm$\,0.008  &   0.086\,$\pm$\,0.008  &   0.094\,$\pm$\,0.011  &   0.089\,$\pm$\,0.011   &   0.088\,$\pm$\,0.005\\  
19 &       ...         &    0.060\,$\pm$\,0.004  &        ...         &   0.067\,$\pm$\,0.006  &   0.068\,$\pm$\,0.006   &   0.065\,$\pm$\,0.003\\ 
25 &  0.082\,$\pm$\,0.007  &    0.083\,$\pm$\,0.007  &   0.085\,$\pm$\,0.007  &   0.088\,$\pm$\,0.008  &   0.087\,$\pm$\,0.008   &   0.086\,$\pm$\,0.004\\  
26 &  0.083\,$\pm$\,0.010  &    0.088\,$\pm$\,0.011  &   0.092\,$\pm$\,0.010  &   0.088\,$\pm$\,0.011  &   0.088\,$\pm$\,0.010   &   0.088\,$\pm$\,0.005\\  
33 &       ...         &    0.058\,$\pm$\,0.006  &        ...         &   0.054\,$\pm$\,0.006  &   0.058\,$\pm$\,0.006   &   0.057\,$\pm$\,0.003\\  
34 &       ...         &    0.093\,$\pm$\,0.007  &        ...         &   0.092\,$\pm$\,0.007  &   0.081\,$\pm$\,0.007   &   0.089\,$\pm$\,0.004\\  
35 &  0.094\,$\pm$\,0.008  &    0.087\,$\pm$\,0.008  &        ...         &   0.084\,$\pm$\,0.010  &   0.087\,$\pm$\,0.010   &   0.087\,$\pm$\,0.005\\  
\hline	
\end{tabular}
\end{minipage}
\end{table*}

The ionic helium abundance relative to hydrogen has been computed from the predicted emissivities for the
\hei\ lines  \citep*{porter07} and for \hb\ \citep{benjamin99} at the corresponding 
\hii\ region electron temperature, as computed in Sect.~\ref{ionic}.

The \hei\ emission lines are affected by underlying stellar absorption, 
which needs to be corrected for. We have followed the approach described by 
\citet{kennicutt03}. %, in which approximate corrections for the \hei\ absorptions are applied.
The corrections  are based  both on direct measurements of the \hei\ EWs for O--type stars 
and  \hei\  EWs predicted by starburst models as a function of age. We have applied here 
the same corrections as \citet{kennicutt03}, except for \hei\ $\lambda$4026 (0.6\,\AA) and  \hei\ $\lambda$4922 (0.3\,\AA), which 
have been estimated from more recent models \citep{rosa05}. These corrections range from $\sim$1\% to 6\% for the 
brightest lines ($\lambda$5876 and $\lambda$6678), $\sim$6-30\% for  $\lambda$4922, and $\sim$2-40\% 
for $\lambda$4471.

\begin{figure}
   \centering
   \includegraphics[bb= 20 8 433 237, clip, width=0.45\textwidth]{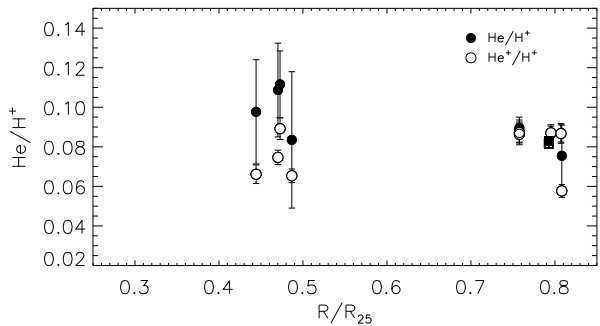}
   \caption{Total  He abundance versus galactocentric distance (black filled circles) for our \hii\ region sample. 
   The ionic He$^+$/H$^+$ abundances are plotted with open circles. %The open triangles correspond to He$^+$/H$^+$ abundances from \citet{Blair82} \hii\ region sample. 
   The total and ionic He abundances of the \hii\ region K932 
   analysed by \citet{esteban09} are marked with a filled and open square, respectively.}
   \label{he_abund2}
\end{figure}

The  He$^+$ abundances computed from the brightest observed \hei\ lines ($\lambda$4026, $\lambda$4471, $\lambda$4922, $\lambda$5876, 
$\lambda$6678) are shown in Table~\ref{abund_he}, together with the weighted average (column~7) of all the individual values. The  He$^+$ abundances have also been computed using the \citet{benjamin99} emissivities. Differences 
in  the abundances derived using the more recent \citet{porter07} emissivities are on average $\sim$2\%, and in all cases below 5\% 
and therefore below the He$^+$/H$^+$ abundance uncertainties.

The total helium abundances have been computed with the analytical relation between the ICF(He) and the softness 
parameter, $\eta$=(O$^+$/O$^{2+}$)/(S$^+$/S$^{2+}$) \citep{vilchez88}, obtained by \citet{Bresolin09} from models 
of \citet*{stasinska01}.
The  ICF(He)  ranges from 1.0 to 1.48 for our \hii\ region sample. 
The resulting total He abundances are shown in  column~7 of Table~\ref{total_abundances} and plotted versus galactocentric radius 
in Fig.~\ref{he_abund2}. The figure includes region K932 \citep{esteban09}, which is in very good agreement with our data for similar
galactocentric radius.
The He abundance appears to be constant within our range of galactocentric distances, with an average value  He/H$^+$\,=\,0.092\,$\pm$\,0.012, 
where the error represents the standard deviation between all regions. 

\section{Oxygen radial abundance gradient}
\label{O_grad}
\begin{figure*}
   \centering
   \includegraphics[bb = 85 340 550 600,clip]{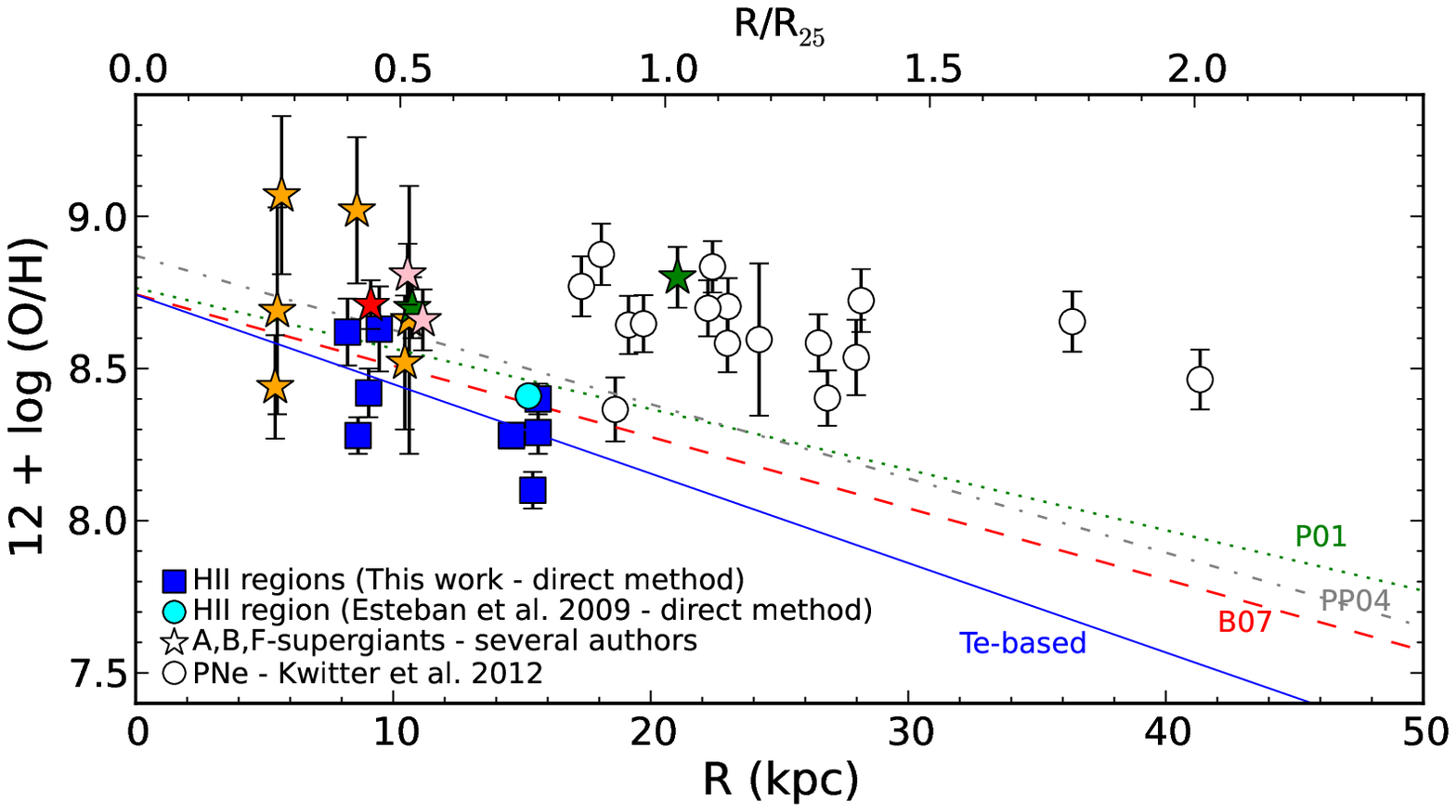}
   \caption{Radial metallicity gradient from the {\em direct} $T_e$--based method obtained from our work 
    (filled blue squares).  A direct abundance measurement from \citet{esteban09} for \hii\ region K932 is indicated 
    with a cyan circle.  A-, B- and F--supergiant oxygen abundances from different authors are indicated with 
    filled yellow \citep{carrie}, green  \citep{Venn}, red \citep{smartt} and pink \citep{Przybilla} stars. 
    We also inlude the recent oxygen abundance determination for M31 PNe by \citet{kwitter}
    with open dots. The solid straight line 
    shows the best linear fit to the \hii\ region {\em direct} oxygen abundance versus galactocentric radius (see Sect.~\ref{O_grad}), while the 
    dotted, dashed and dot--dashed straight lines show the best linear fits to the oxygen abundances obtained from the \citet{P01}, \citet{b07} and  
    \citet{PP04} strong--line methods, respectively, applied to the whole sample of \hii\ regions  available in the literature (see Sect.~\ref{strong}).}
   \label{O_abund}
\end{figure*}

%\subsection{Direct measurements}
The \hii\ region oxygen abundances in M31 obtained in this paper from the direct $T_e$--based method are plotted
as a function of galactocentric radius in Fig.~\ref{O_abund} (blue squares). The linear regression to this data plus the \hii\ region K932 by \citet{esteban09} yields the following solution:
\begin{equation}
\label{fitOH}
\mbox{12+log(O/H)}=  8.72(\pm0.18)-0.028(\pm0.014)\times\mbox{R (kpc)} 
\end{equation}
or equivalently,
\begin{equation}
\mbox{12+log(O/H)}= 8.72(\pm0.18) -0.56 (\pm0.28)\times\mbox{(R/R$_{25}$)}
\end{equation}
%Coef. correlacion ~0.6
%Our data then indicates the presence of a oxygen abundance gradient across the disc of M31.
The slope that we obtain, $-0.028\pm0.014$~dex~kpc\me, represents a rather shallow oxygen abundance gradient in M31. The central abundance is only slightly above solar \citep[12 +$\log$(O/H)$_\odot$=8.69, ][]{asplund}.
We are aware that our data are virtually representative of only two distinct galactocentric distances, and further 
data at different  distances from the M31 center would be desirable to better determine the
abundance gradient from the $T_e$--based method  alone.

This is the first time that an oxygen abundance gradient from auroral lines is obtained for M31.
Previous attempts to measure the oxygen abundance gradient in this galaxy from \hii\ regions were performed 
from  diagnostics based on strong emission lines, because of the lack  of the faint auroral lines necessary to measure the 
electron temperatures.
The early papers by \cite{dennefeld81} and \cite{Blair82}
contain \hii\ region samples of similar sizes (8 and 11 regions, respectively) and 
make use of  empirical calibrations of the [\oiii]/[\nii] line ratio \citep{alloin} and the R$_{23}$ parameter 
(R$_{23}$=([\oiii] + [\oii])/H$\beta$, \citealt{pagel79}) to derive the $T_e$ from which the O/H is estimated. 
Their O/H vs. galactocentric radius relations have a lot of scatter,
but are in agreement with each other and seem compatible with super--solar oxygen abundance in the M31 center 
(12+log(O/H) $>$ 9.1) and an abundance  gradient with a slope of approximately  $-$0.03~dex~kpc (from  Fig.~4 and Fig.~6. of \citet{dennefeld81}
and \citet{Blair82}, respectively). This value is in agreement with our slope, but their metallicities are about
0.5~dex systematically higher than our determinations. 

Additional authors have recalculated the oxygen abundance gradient in M31, by applying different 
calibrations of the R$_{23}$ parameter to the data obtained by \citet{dennefeld81} and \citet{Blair82}. 
\citet{vila-costas} used the R$_{23}$ calibration by \citet{edmunds_pagel}, with adjustments at high and
low metallicities \citep{edmunds,skillman}. They obtained a somewhat steeper slope $-$0.043~dex~kpc\me.
In contrast, \citet{zaritsky94}, using the same dataset, obtained a shallower abundance gradient  
($-0.018\pm0.006$~dex~kpc\me) with a combination of three different calibrations of the R$_{23}$ parameter 
\citep*{edmunds_pagel,dopita_evans,McCall}.  In both cases a super--solar central abundances (again 12+log(O/H) $>$ 9.1) was obtained.

The most recent spectroscopic observations of a sample of \hii\ regions in M31 were performed by \citet{Galarza}. Their sample
comprises 46 \hii\ regions which the authors classify according to their morphological type. These authors inferred a 
metallicity gradient of $-0.06 \pm 0.03$~dex~kpc\me\ from the dependence of the R$_{23}$ parameter of the regions
classified as {\em center--brightened} ($\sim$20 \hii\ regions) with the radial distance from the galaxy center.
This gradient is steeper compared to our and previous determinations of the oxygen abundance 
gradient of M31, but  given its large uncertainty this result is still compatible with the shallower slope we find.
Further spectroscopic observations of six  \hii\ regions in M31 were obtained by \citet{B99}, but 
for a different purpose, and no abundance estimations were performed. 

More recent published values of the M31 oxygen abundance gradient from \hii\ regions come from the reanalysis 
of the datasets described above \citep{dennefeld81,Blair82,Galarza} using different empirical calibrations
(\citealt*{pagel80}; \citealt{M91}; \citealt{zaritsky94}; \citealt{P01}). Depending on the empirical calibration applied,
the slope ranges from $-0.013$ to $-0.027$~dex~kpc\me, and the central abundance varies between 8.70 and 9.20~dex 
\citep{smartt,carrie}.

\subsection{Strong-line methods}
\label{strong}
The previous section makes it clear that  the usage of different  calibrations of ratios of bright nebular 
emission lines ({\em strong-line methods}), which are readily accessible observationally to determine 
the oxygen abundance gradient from a given data set, yields systematic 
and non-negligible differences on both  the metallicity gradient and on the central (R=0~kpc) 
oxygen abundance. This is well known from the detailed analysis of different galaxies from many authors
 \citep[e.g.][]{montero05,kewley08,Bresolin09,b09m83,LopezEsteban}.
In particular, strong-line methods calibrated via grids of photoionization models generally provide 
higher metallicities than those calibrated empirically via sets of \hii\ region 
oxygen abundance determinations obtained from $T_e$ estimates from detected auroral lines.

Given the still limited number of regions for which we have a direct determination of the electron
temperature in M31, it is useful to apply different bright--line metallicity indicators to the whole set of M31
\hii\ regions observed so far to enlarge the range in galactocentric distance and the number of regions for a better statistical significance
of the gradient parameters.
As a first step, we have applied strong--line methods to the whole set of 31 \hii\ regions analyzed in this paper
(those with and without detections of auroral lines), limiting our choice to six widely used methods: (a) The R$_{23}$ method, 
as calibrated empirically by \citet[][P01]{P01} and \citet[][PT05]{PT05} from spectroscopic data of \hii\ regions with measured T$_e$--based oxygen abundances.  
(b)   The R$_{23}$ method, as calibrated from theoretical model grids by  \citet[][M91]{M91}, through the \citet*{kuzio} parametrization, which takes into account the 
ionization parameter of the \hii\ regions.  (c) The N2O2 method, i.e. the empirical calibration of the [\nii]$\lambda$6583/[\oii]$\lambda$3727 line ratio
by \citet[][B07]{b07} with extragalactic \hii\ regions with $T_e$--based oxygen abundances, and the calibration obtained by \citet[][]{KD02} from stellar population synthesis 
and photoionization models. (d) The N2\,$\equiv\log$([\nii]$\lambda$6583/\ha) method, empirically calibrated by \citet[][PP04]{PP04} with 
\hii\ regions with $T_e$--based oxygen abundances or detailed photoionization modelling.
\begin{figure*}
   \centering
   \includegraphics[bb = 30 180 570 530,clip, width=0.95\textwidth]{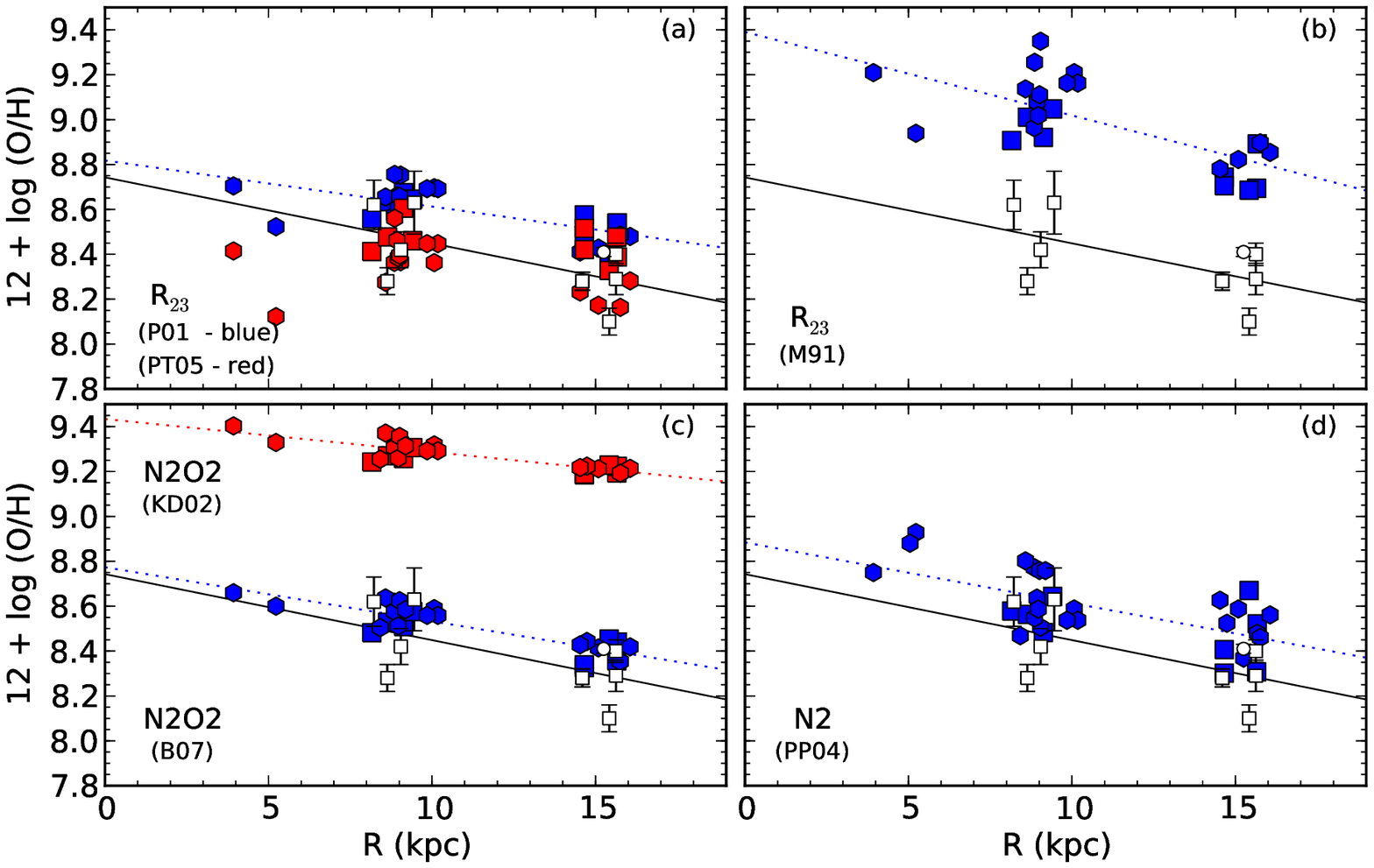}
  \caption{Radial oxygen abundance gradient in M31 from different calibrations of strong-line ratios applied to the \hii\ regions analyzed in this paper (filled symbols):  {\bf (a)} \citet[][P01]{P01} and \citet[][PT05]{PT05} empirical calibrations of the  R$_{23}$ indicator.  {\bf (b)} R$_{23}$ strong--line method as calibrated by \citet{M91}.   {\bf (c)} N2O2 (=[\nii]/[\oii]) as calibrated by \citet[][B07]{b07} and \citet[][KD02]{KD02} {\bf (d)} N2 =log([\nii]6583/\ha) as empirically calibrated by  \citet[][PP04]{PP04}.    The dotted straight lines show the best linear fit to the oxygen abundance gradient obtained from the corresponding strong--line methods. The  fit parameters are shown in Table~\ref{fits_empirical}. The solid straight line shows the best fit to the oxygen abundances obtained from direct measurements of $T_e$ (Eq.~\ref{fitOH}). Filled squares indicate \hii\ regions with detections of auroral lines, while filled hexagons indicate regions with no detections of auroral lines. The open symbols show the direct $T_e$--based oxygen abundances derived in this paper.}
\label{empirical_our_regions}
\end{figure*}

The results from the application of these methods to our \hii\ region sample are plotted in Fig.~\ref{empirical_our_regions}
as a function of the galactocentric radius.
For the three R$_{23}$-based calibrations we have used the upper branch, as indicated by the discriminator 
 [\nii]$\lambda$6583/[\oii]$\lambda$3727 \citep{kuzio,kewley08}.  
For comparison we also plot (with open symbols) the $T_e$--based oxygen abundances derived in this paper 
together with the best linear fit to the data (solid line).
The oxygen abundances obtained with the R$_{23}$ parameter as calibrated by \citet{M91} and the ones obtained 
with the N2O2 calibration of \citet{KD02} are $\sim$0.4-0.7~dex larger than the
ones obtained from the other methods, as expected when comparing  empirical and theoretical model grid calibrations. 
Table~\ref{fits_empirical} shows the best linear fit parameters for the oxygen gradient obtained from the different strong--line methods.

It is apparent from  Fig.~\ref{empirical_our_regions} and Table~\ref{fits_empirical} that the scatter varies between the different O/H indicators, being larger for
the R$_{23}$--based methods and N2, and smaller in the case of N2O2  which presents 
the highest correlation coefficient (around $-0.85$ for both calibrations of N2O2,  with a root mean squared ({\em rms}) residual scatter 
from the fit of 0.05~dex and 0.03~dex for the \citet{b07} and \citet{KD02} calibrations, respectively
(cf. 0.12~dex for N2 and for the PT05 calibrations and 0.14~dex for  R$_{23}$ as calibrated by \citealt{M91}).

Fig.~\ref{empirical_our_regions} and Table~\ref{fits_empirical} show  that the slope of the oxygen abundance gradient 
obtained from the \citet{P01} calibration of the R$_{23}$ parameter, from  N2, and  
from  N202 as calibrated by \citet{b07}  agree within the errors, and also agree with the oxygen abundance gradient obtained from 
the direct $T_e$--based oxygen abundances obtained in this paper. The \citet{M91} calibration of the R$_{23}$ parameter agrees 
marginally with the N2 calibration and with the  $T_e$--based oxygen gradient.
The N202 method as calibrated by \citet{KD02} and the \citet{PT05} calibration give shallower oxygen abundance gradients, but we note that
the correlation of the abundances from the \citet{PT05} method with galactocentric radius is poor (correlation coefficient $\sim -0.2$).

\begin{figure*}
   \centering
   \includegraphics[bb = 30 180 570 530,clip, width=0.95\textwidth]{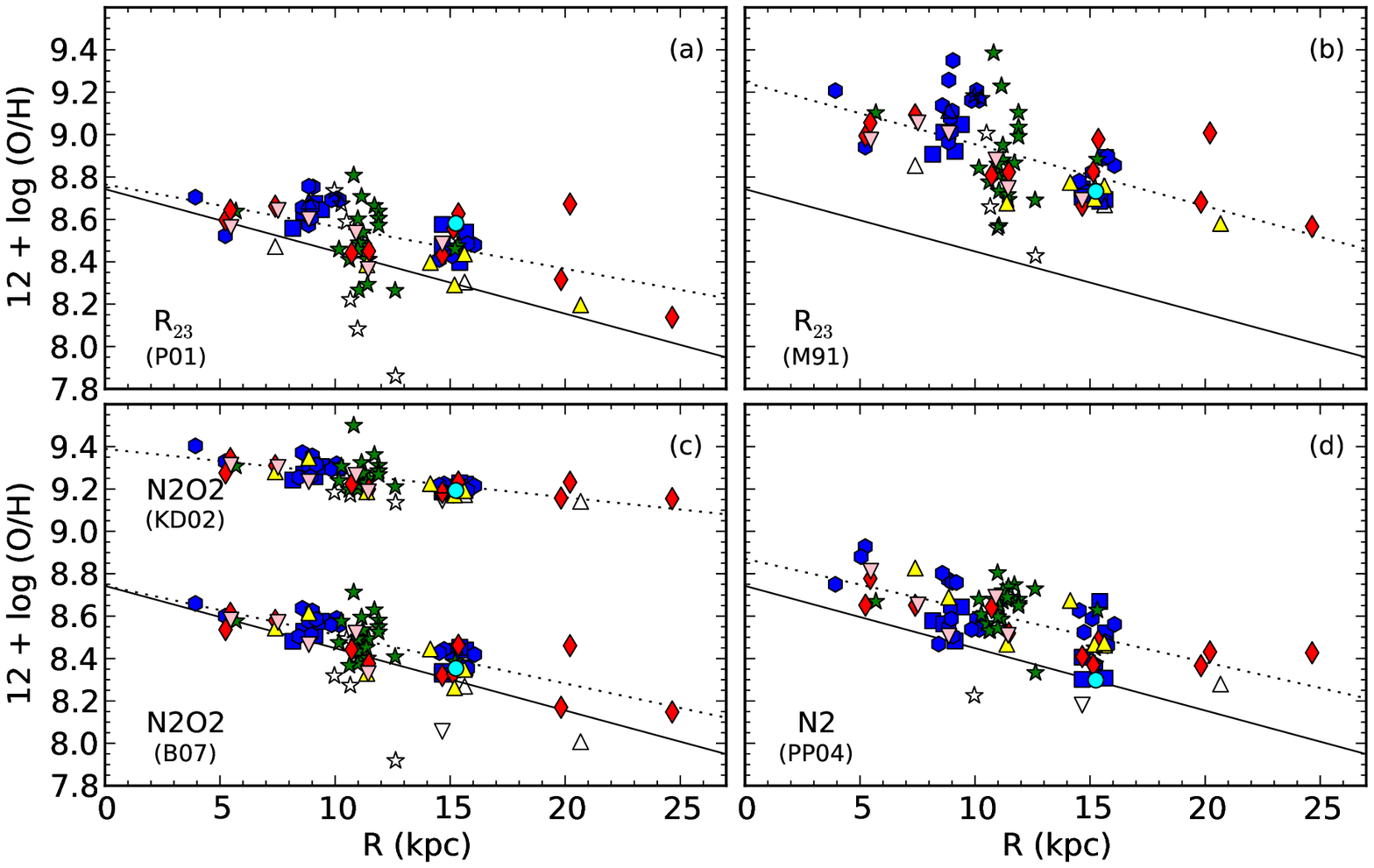}
    \caption{Same as Fig.~\ref{empirical_our_regions}, but for the \hii\ regions analyzed in this paper plus
    \hii\ regions spectroscopically observed by other authors: \citet{dennefeld81}, yellow triangles;  \citet{Blair82}, red diamonds; 
    \hii\ regions classified as 'center--bright' by \citet{Galarza}, green stars; \citet{B99}, pink inverted triangles; 
     region K932 (cyan circle) observed by \citet{esteban09},  and \hii\ regions analyzed in this paper (blue squares and 
     hexagons, as in Fig.~\ref{empirical_our_regions}). Open symbols indicate \hii\ regions with uncertain values in the relevant 
     line ratios for a given empirical calibration, as stated by the corresponding authors.
     The dotted straight lines show the best linear fit to the oxygen abundance gradient obtained from the strong--line methods applied 
     to the whole \hii\  region dataset (excluding regions with uncertain line ratios).
     The solid straight line shows the oxygen abundance gradient obtained from the $T_e$--based method (Eq.~\ref{fitOH}). }
   \label{empirical_all_authors}
\end{figure*}

The six strong--line methods have also been applied to all spectroscopic data available 
in the literature for  \hii\ regions in M31  \citep{dennefeld81,Blair82,Galarza,B99,esteban09}. The galactocentric distances have
been recalculated from their corresponding equatorial coordinates (as reported by the  various authors) and 
using the same  galactic parameters as for the estimation of the galactocentric distances of our \hii\ region sample (see Table~\ref{M31 params}).
Only the \hii\ regions
morphologically classified as {\em center--brightened}  from the \citet{Galarza} sample have been
considered. We also note that these authors and \citet{B99} only provide the total [\nii] ($\lambda$6548+$\lambda$6583) emission. In order 
to apply the N2 and the N2O2 methods to these samples, we have estimated the
expected [\nii]$\lambda$6583 emission by assuming [\nii]$\lambda$6583 = 3$\times$[\nii]$\lambda$6548.
Also, in the cases in which [\oiii]$\lambda$4959 is not detected, we assumed that [\oiii]$\lambda$$\lambda$4959+5007=(4/3) [\oiii]$\lambda$5007.
The full \hii\ region sample comprises 85 objects.
\begin{table*}
\centering
\begin{minipage}{120mm}
\centering
\caption{Linear least--squares fit parameters to the oxygen abundance distribution across the M31 disk obtained from different 
strong--line methods (see Sect.~\ref{strong}) applied to our \hii\ region sample and to the whole set of spectroscopic data on M31 \hii\ 
regions published so far: this paper, \citet{dennefeld81}, \citet{Blair82}, \citet{Galarza}, \citet{B99}, \citet{esteban09}. }
\label{fits_empirical}
\begin{tabular}{l c  c c c c}     % 6 columns
\hline
Method                              &  slope            &  zero point  &   Correlation&  No.           &     Scatter\footnote{{\em rms} residual scatter, estimated as the square root of the arithmetic mean value of the squares of the difference between fitted and measured data.} \\
                                    &  (dex~kpc\me)     &    (dex)     & coefficient  &  data points    &     (dex)  \\
\hline
\multicolumn{6}{c}{Our \hii\ region sample} \\
\hline
R$_{23}$ (P01)                        & $-$0.021\,$\pm$\,0.004 & $8.82\pm0.05$ &  $-$0.69      &  25   &   0.08    \\
R$_{23}$ (PT05)                       & $-$0.007\,$\pm$\,0.007 & $8.45\pm0.08$ &  $-$0.20      &  25   &   0.12    \\
R$_{23}$ (M91)                        & $-$0.037\,$\pm$\,0.008 & $9.39\pm0.09$ &  $-$0.70      &  25   &   0.14    \\
N2O2 (B07)                           & $-$0.024\,$\pm$\,0.003 & $8.77\pm0.03$ &  $-$0.86      &  28   &   0.05    \\
N2O2 (KD02)                          & $-$0.015\,$\pm$\,0.002 & $9.43\pm0.02$ &  $-$0.84      &  28   &   0.03    \\
N2 (PP04)                            & $-$0.027\,$\pm$\,0.006 & $8.88\pm0.07$ &  $-$0.65      &  31   &   0.12    \\
\hline
\multicolumn{6}{c}{\hii\ regions from all authors} \\
\hline
R$_{23}$ (P01)                        &  $-$0.020\,$\pm$\,0.004 & $8.76\pm0.05$ &  $-$0.55      &  70   &   0.12    \\
R$_{23}$ (PT05)                       &  $-$0.009\,$\pm$\,0.005 & $8.45\pm0.06$ &  $-$0.24      &  70   &   0.15    \\
R$_{23}$ (M91)                        &  $-$0.029\,$\pm$\,0.005 & $9.25\pm0.06$ &  $-$0.59      &  70   &   0.15    \\
N2O2 (B07)                           &  $-$0.023\,$\pm$\,0.002 & $8.74\pm0.03$ &  $-$0.75      &  75   &   0.08    \\
N2O2 (KD02)                          &  $-$0.011\,$\pm$\,0.002 & $9.39\pm0.02$ &  $-$0.66      &  75   &   0.05    \\
N2 (PP04)                            &  $-$0.024\,$\pm$\,0.003 & $8.87\pm0.04$ &  $-$0.64      &  82   &   0.11    \\
\hline
\end{tabular}
\end{minipage}
\end{table*}

The results are plotted in Fig.~\ref{empirical_all_authors} and the corresponding least--square fits parameter shown in 
Table~\ref{fits_empirical}. We obtain a similar behaviour as for  our data alone: we observe an agreement between the
$T_e$--based oxygen gradient and the slopes obtained from strong-line methods with the exception of the \citet{PT05} empirical 
calibration and the N2O2 as calibrated by \Citet{KD02}. These methods yield a substantially shallower slope ($-0.009\pm0.005$ and 
$-0.011\pm0.002$~dex~kpc\me, respectively). In the case of the \citet{PT05} calibration, we find again that the correlation is poor 
(correlation coefficient $-0.24$). 
This lack of agreement of the \citet{PT05} calibration with the  $T_e$--based oxygen abundances is surprising and it is
in contradiction with \citet{LopezEsteban}, who conclude that the \citet{PT05} calibration is nowadays the best empirical
method to estimate oxygen abundances when auroral lines are not detected.

We should note that in all empirical calibrations we slightly improve the correlation coefficient and the scatter when the 
\citet{Galarza} sample is not taken into account, but the slope and intercept of the fits remain unchanged.

We have computed the weighted average of the slope of the oxygen abundance gradient from the strong-line methods with lower scatter and 
with a better match with our direct  $T_e$--based oxygen abundances (i.e. with the N2O2 method as calibrated by \citealt{b07}, 
the N2 calibrated by \citealt{PP04}, and the \citealt{P01} calibration) with weights equal to the 
reciprocal of the  errors, yielding the value
\begin{equation} 
d\,log(O/H)/dR = -0.023\pm0.002\mbox{~dex~kpc\me}
\end{equation}
that we adopt  as the slope of the oxygen abundance gradient in M31.

\section{Comparison with other metallicity indicators}
\label{comparison}

We have compared the \hii\ region abundance gradient with the results obtained from young, BA-type supergiants and from planetary nebulae, which trace an older stellar population. 
Despite the relatively small distance to the Andromeda galaxy and the importance of this extragalactic neighbour in the context of galactic
evolution studies, the analysis of a sizeable sample of blue supergiant spectra for the determination of stellar metallicities is still lacking.
The situation has remained virtually unchanged during the past decade, since the publication of stellar metallicities for 7 B-type supergiants
 by \citet{carrie}, and, somewhat paradoxically, lags behind the quantitative spectroscopic study of massive stars in considerably more distant 
galaxies \citep[e.g.~][]{kud08,kud12}. One of these stars had been originally analyzed by \citet{smartt}.
Additionally,  two A- and one F-type supergiants were investigated by \citet{Venn}. More recently \citet*{Przybilla}
revisited the metallicity derived for one of the A-type stars with more accurate non-LTE calculations, and analyzed a new  A-type star. It should be pointed out that 
the B-type metallicities obtained by \citet{carrie} were based on an LTE analysis, except for one case, in which the non-LTE 
metallicities were found to be consistent, within the relatively large (between $\sim$0.2 and 0.4 dex) errors, even though they 
are 0.07~dex lower in the case of oxygen (a similar result is obtained comparing the LTE vs. non-LTE analysis of the star in common with \citealt{smartt}).

These studies highlighted the difficulty of comparing the stellar and the nebular metallicities, due to the systematic differences 
in \hii\ region abundances obtained from different strong--line methods. For this reason, the results of the comparison were inconclusive.
Our new study of \hii\ regions in M31 was mostly motivated by the need to acquire {\em direct} $T_e$--based nebular abundance. As this 
paper shows, we have  partially succeeded in our goal, but the detection of auroral lines in \hii\ region spectra remains elusive for M31 
even with 10m-class telescopes.

In Fig.~\ref{O_abund} we include the O/H abundances derived for the B-type supergiant stars by Trundle et al.~(2002, yellow star symbols) and Smartt et al.~(2001, red star symbol), and for the AF stars studied by Venn et al.~(2000, green star symbols) and Przybilla et al.~(2006, pink star symbols). When available, only the non-LTE result is shown.  Galactocentric distances have been recalculated from our adopted M31 parameters (see Table~\ref{M31 params}). The comparison is somewhat complicated by the large errors in the B supergiant abundances, but it is obvious that a systematic offset is present in relation to the {\em direct} abundances we determined from \hii\ regions. If we focus on the A-type supergiants (galactocentric distance  $\sim$11.7~kpc), which have more accurate abundances, the nebular O/H ratio lies $\sim$0.3~dex below the stellar value. On the other hand, among the gradients obtained from strong-line diagnostics, we find that
the N2 method, as calibrated by \citet{PP04}, is in acceptable agreement with the absolute value provided by the stars (the difference relative to the stars is reduced to $\sim$0.1 dex). This situation contrasts the results obtained
by our group in other, lower metallicity galaxies, for example NGC~300, where excellent agreement is found between the metallicities obtained for stars and \hii\ regions using {\em both} the direct $T_e$--based method and N2 \citep{Bresolin09}. A situation similar to what we encounter in M31 is  also found for M81 \citep{patterson, kud12}. A possible interpretation of this discrepancies will be provided in the next section.

Recently \citet{kwitter} published the abundance analysis of 16 PNe from [\oiii]$\lambda$4363 detections in the outer disk ($R>20$~kpc) of M31. The radial distribution
of the O/H ratio is included in Fig.~\ref{O_abund} (open circles; we recalculated the galactocentric distances from our adopted orientation parameters, Table~\ref{M31 params}, finding only minor differences). We note a striking offset between \hii\ regions and PNe direct abundances (although we note that their target 16 falls near the \hii\ region locus). It is not clear whether the two samples of objects are directly comparable, however, due to the fact that the PNe belong to the outer disk of the galaxy. Difficulties in the interpretation of the offset might also arise from the complicated chemical history of M31 (e.g.~Davidge et al.~2012 for a recent summary). Finally, we note that the slope of the PNe O/H gradient obtained by \citet{kwitter},
$-0.011\pm0.004$ dex\,kpc$^{-1}$,  is somewhat flatter than the one we derive from \hii\ regions. 
We add that a flatter gradient for PNe compared to \hii\ regions  has also been recently determined  in the galaxy NGC~300 (Stasinska et al. 2012, in preparation), and corroborates recent model predictions of galactic chemical evolution \citep{pilkington}.

\section{Discussion and conclusions}
\label{discussion}

One of the main conclusions obtained in this paper is that the slope of the oxygen abundance gradient in M31 from the analysis of the \hii\ regions is quite robust, 
approximately $-0.023$~dex\,kpc$^{-1}$, regardless of the choice of abundance diagnostic used (with the notable exception of the \citealt{PT05} and \citealt{KD02} calibrations). 
However, the $\sim$0.3 dex offset between the {\em direct} abundances (obtained from the auroral line detections) and the stellar abundances is somewhat disturbing, especially considering the fact that the direct abundances also lie significantly below those obtained from strong-line methods that, by construction, should be in agreement with the direct determinations. 
Since the discrepancy could be a result of the higher metallicity of M31 compared to other galaxies where such offset is not observed,
we suggest an interpretation  based on biases affecting the direct abundances. As is well known, with increasing O/H the detection of the auroral lines in the spectra of \hii\ regions becomes increasingly difficult. This is due to the exponential dependence of the line emissivities on the electron temperature, and the decreasing nebular temperatures with increasing metallicity due to more efficient cooling.
As a result, very few auroral line detections exist in the literature for extragalactic \hii\ regions with metallicity above solar. Therefore, given the metallicity (around the solar value) found for the supergiant stars in M31 it is not surprising that we only have a few [\oiii]$\lambda$4363 detections. Under these conditions we would preferentially  detect this line just for some of the hottest (less metal enriched) \hii\ regions, which leads to a bias in the derived average oxygen abundance: the abundance estimated from the few [\oiii]$\lambda$4363 detections would underestimate the mean O/H ratio. This would explain why the auroral line-based abundances lie below the strong line abundances obtained from methods such as N2 and N2O2, which have been calibrated using \hii\ regions with detections of auroral lines.
If we assume an observational abundance scatter of $\sim0.07$~dex (comparable to what has been measured in other nearby galaxies, see \citealt{Bresolin09}, \citealt{kennicutt03}, \citealt{bresolin11}), we can expect to systematically underestimate the oxygen abundance by as
much as $\sim$0.15~dex if we assume that our auroral line detections refer to the hottest nebulae only, i.e.~are not representative of the mean \hii\ region population. Most of the remaining discrepancy with the stellar abundances could be explained by depletion of oxygen 
onto dust grains, which has been estimated to be $\sim$0.12~dex at high metallicity (e.g.~\citealt{mesa-delgado09}, \citealt{peimbert2}).
If we account for these effects, the central abundance of M31 would be approximately 12+log(O/H) = 8.94\,$\pm$\,0.10~dex; i.e.~between 1.4 and 2.2 times the solar value. 

Finally, we  also  point out that a discrepancy between {\em direct} nebular abundances and stellar abundances increasing with metallicity would naturally result from
the presence of inhomogeneities in the temperature distribution within an \hii\ region \citep{peimbert07} or by
 the effect recently reported by \citet*{nicholls12}, who proposed that the energy distribution of 
electrons in \hii\ regions deviates from a Maxwell-Boltzmann one.
Further tests of the mechanisms proposed here to explain the systematic offset between $T_e$-based \hii\ region abundances and
stellar metallicities would require to increase the sample of \hii\ regions with {\em direct} abundance determinations for metal-rich (super-solar) galaxies,
together with a program aimed at measuring  metallicities of blue supergiant stars. 
In the specific case of M31, in particular, more accurate stellar abundance data for a large sample of B-type stars would be highly desirable.

\section*{Acknowledgments}
We kindly acknowledge Carrie Trundle for her contribution in the initial phases of this project.  We also thank the anonymous
referee for an encouraging report and constructive comments.
A. Zurita acknowledges support from the Spanish ``Plan Nacional del Espacio
de Ministerio de Educaci\'on y Ciencia'' (via grant C-CONSOLIDER AYA
2007-67625-C02-02) and from the  ``Junta Andaluc\'\i a'' local government through the FQM--108 project
and  through the ``ayudas individuales'' program.
F. Bresolin gratefully acknowledges the support from the National Science 
Foundation grants AST-0707911 and AST-1008798.
A. Zurita would like to thank the Institute for Astronomy at Manoa, Honolulu for hospitality. F. Bresolin is grateful 
for the kind hospitality received at the Departamento de F{\'i}sica Te\'orica y del Cosmos in Granada.

\label{lastpage}

\end{document}